\documentclass[12pt,preprint]{aastex}

\shorttitle{The Cosmic Microwave Background}
\shortauthors{Kosowsky}

\begin{document}

\title{The Cosmic Microwave Background}

\author{Arthur Kosowsky}
\affil{Department of Physics and Astronomy, Rutgers University,
    Piscataway, NJ 08854-8019}
\email{kosowsky@physics.rutgers.edu}

\begin{abstract}
This set of lectures provides an overview of the 
basic theory and phenomenology of the cosmic microwave background. 
Topics include a brief historical review; the physics of
temperature and polarization fluctuations; acoustic oscillations of
the primordial plasma; the space of inflationary cosmological
models; current and potential constraints on these models from
the microwave background; and constraints on inflation.
These lectures were given at the Italian Society of Gravitational
Physics Summer School ``Relativistic Cosmology: 
Theory and Observation'' in Como, Italy (May 2000).
\end{abstract}

\section{Introduction}

It is widely accepted that the field of cosmology is entering an era
dubbed ``precision cosmology.'' Data
directly relevant to the properties and evolution of the universe is
flooding in by the terabyte (or soon will be). Such vast quantities of
data were the purview only of high energy physics just a few years
ago; now expertise from this area is being coopted by some astromers
to help deal with our wealth of information. In the past decade,
cosmology has gone from a data-starved science in which often highly
speculative theories went unconstrained to a data-driven pursuit where
many models have been ruled out and the remaining ``standard
cosmology'' will be tested with stringent precision.

The cosmic microwave background radiation is at the center of this
revolution. The radiation present
today as a 2.7 K thermal background originated when the universe was
denser by a factor of $10^9$ and younger by a factor of around
$5\times 10^4$. The radiation provides the most distant direct image
of the universe we can hope to see, at least until gravitational
radiation becomes a useful astronomical data source. The microwave
background radiation is extremely uniform, varying in temperature by
only a few parts in $10^5$ over the sky (apart from an overall dipole
variation arising from our peculiar motion through the microwave
background's rest frame); its departure from a perfect blackbody
spectrum has yet to be detected. 

The very existence of the microwave background provides crucial
support for the Hot Big Bang cosmological model: the universe began in
a very hot, dense state from which it expanded and cooled.  
The microwave background visible today was once in thermal
equilibrium with the primordial plasma of the universe, and the
universe at that time was highly uniform. Crucially, the universe
could not have been perfectly uniform at that time or no structures
would have formed subsequently. The study of small temperature and
polarization fluctuations in the microwave background, reflecting
small variations in density and velocity in the early universe, have
the potential to provide the most precise constraints on the overall
properties of the universe of any data source. The reasons are that
(1) the universe was very simple at the time imaged by the microwave
background and is extremely well-described by linear perturbation
theory around a completely homogeneous and isotropic cosmological
spacetime; and (2) the physical processes relevant at that time are
all simple and very well understood. The microwave background is
essentially unique among astrophysical systems in these regards.

The goal behind these lectures is to provide a qualitative description
of the physics of the microwave background, an appreciation for the
microwave background's cosmological importance, 
and an understanding of what kinds of
constraints may be placed on cosmological models. These lectures are
not intended to be a definitive technical reference to the microwave
background.  Unfortunately, such a reference does not really exist at
this time, but I have attempted to provide pedagogically useful
references to other literature. I have also not attempted to give a
complete bibliography; please do not consider this article to give
definitive references to any topics mentioned. A recent review of the
microwave background with a focus on potential particle physics
constraints is Kamionkowski and Kosowsky (1999). A more general review
of the microwave background and large-scale structure
with references to many early microwave background articles is
White {\it et al.}\ (1994).

\section{A Brief Historical Perspective}
\label{history}

The story of the serendipidous discovery of the microwave background
in 1965 is widely known, so I will only briefly summarize it here.  A
recent book by the historian of science Helge Kraugh (1996) is a
careful and authoritative reference on the history of cosmology, from
which much of the information in this section was obtained.  Arno
Penzias and Robert Wilson, two radio astronomers at Bell Labs in
Crawford, New Jersey, were using a sensitive microwave horn radiometer
originally intended for talking to the early Telstar
telecommunications satellites.  When Bell Labs decided to get out of
the communications satellite business in 1963, Penzias and Wilson
began to use the radiometer to measure radio emission from the
Cassiopeia A supernova remnant.  They detected a uniform noise source,
which was assumed to come from the apparatus. But after many months of
checking the antenna and the electronics (including removal of a
bird's nest from the horn), they gradually concluded that the signal
might actually be coming from the sky. When they heard about a talk
given by P.J.E. Peebles of Princeton predicting a 10 K blackbody
cosmological background, they got in touch with the group at Princeton
and realized they had detected the cosmological radiation.  At the
time, Peebles was collaborating with Dicke, Roll, and Wilkinson in a
concerted effort to detect the microwave background. The Princeton
group wound up confirming the Bell Labs discovery a few months
later. Penzias and Wilson published their result in a brief paper with
the unassuming title of ``A measurement of excess antenna temperature
at $\lambda = 7.3$ cm'' (Penzias and Wilson 1965); a companion paper
by the Princeton group explained the cosmological significance of the
measurement (Dicke et al 1965). The microwave background detection
was a stunning success of the Hot Big Bang model, which to that point
had been well outside the mainstream of theoretical physics.  The
following years saw an explosion of work related to the Big Bang model
of the expanding universe. To the
best of my knowledge, the Penzias and Wilson paper was the
second-shortest ever to garner a Nobel Prize, awarded in 1978. (Watson
and Crick's renowned double helix paper wins by a few lines.)

Less well known is the history of earlier probable detections of
the microwave background which were not recognized as such.
Tolman's classic monograph on thermodynamics in an expanding universe
was written in 1934, but a blackbody relic of the early universe
was not predicted theoretically until 1948 by Alpher
and Herman, a by-product of their pioneering work on nucleosynthesis
in the early universe. Prior to this, Andrew McKellar (1940) had observed
the population of excited rotational states of CN molecules in 
interstellar absorption lines, concluding that it was consistent with being
in thermal equilibrium with a temperature of around 2.3 Kelvin. 
Walter Adams also made similar measurements (1941). Its
significance was unappreciated and the result essentially forgotten,
possibly because World War II had begun to divert much of the world's
physics talent towards military problems. 

Alpher and Herman's prediction of a 5 Kelvin background
contained no suggestion of its detectability with available technology
and had little impact. Over the next decade, George Gamow and
collaborators, including Alpher and Herman, made a variety of
estimates of the background temperature which fluctuated 
between 3 and 50 Kelvin (e.g. Gamow 1956). 
This lack of a definitive temperature might
have contributed to an impression that the prediction was less
certain than it actually was, because it aroused little interest
among experimenters even though microwave technology had been
highly developed through radar work during the war. At the same
time, the incipient field of radio astronomy was getting started.
In 1955, Emile Le Roux undertook an all-sky survey at a wavelength
of $\lambda = 33$ cm, finding an isotropic emission corresponding
to a blackbody temperature of $T=3\pm 2$ K (Denisse {\it et al.} 1957). This was
almost certainly a detection of the microwave background, but its
significance was unrealized. Two years later, T.A. Shmaonov
observed a signal at $\lambda=3.2$ cm corresponding to a blackbody
temperature of $4\pm 3$ K independent of direction (see Sharov and
Novikov 1993, p.~148).  The significance
of this measurement was not realized, amazingly, until 1983! 
(Kragh 1996). Finally in the early 1960's the pieces began to
fall in place: Doroshkevich and Novikov (1964) emphasized the detectability
of a microwave blackbody as a basic test of Gamow's Hot Big Bang
model. Simultaneously, Dicke and
collaborators began searching for the radiation, prompted by
Dicke's investigations of the physical consequences of the
Brans-Dicke theory of gravitation. They were soon scooped by
Penzias and Wilson's discovery.

As soon as the microwave background was discovered, theorists
quickly realized that fluctuations in its temperature would have
fundamental significance as a reflection of the initial perturbations
which grew into galaxies and clusters. Initial estimates
of the amplitude of temperature fluctuations were a part in a hundred;
this level of sensitivity was attained by experimenters after a few
years with no observed fluctuations. Thus began a quarter-century
chase after temperature anisotropies in which the
theorists continually revised their estimates of the fluctuation
amplitude downwards, staying one step ahead of the experimenters'
increasingly stringent upper limits. Once the temperature fluctuations
were shown to be less than a part in a thousand, baryonic density
fluctuations did not have time to evolve freely into the nonlinear
structures visible today, so theorists invoked a gravitationally dominant
dark matter component (structure formation remains one of the
strongest arguments in favor of non-baryonic dark matter). By the
end of the 1980's, limits on temperature fluctuations were well below
a part in $10^4$ and theorists scrambled to reconcile
standard cosmology with this small level of primordial fluctuations.
Ideas like late-time phase transitions at redshifts less than 
$z=1000$ were taken seriously as a possible way to evade the
microwave background limits (see, e.g., Jaffe {\it et al.} 1990). Finally, the
COBE satellite detected fluctuations at the level of a few parts
in $10^5$ (Smoot {\it et al.} 1990), just consistent with structure formation in
inflation-motivated Cold Dark Matter cosmological models. The COBE
results were soon confirmed by numerous ground-based and balloon
measurements, sparking the intense theoretical and experimental
interest in the microwave background over the past decade.

\section{Physics of Temperature Fluctuations}
\label{physics}

The minute temperature fluctuations present in the microwave
background contain a wealth of information about the fundamental
properties of the universe.  In order to understand the reasons for
this and the kinds of information available, an appreciation of the
underlying physical processes generating temperature and polarization
fluctuations is required. This section and the following one give a
general description of all basic physics processes involved in
producing microwave background fluctuations. 

First, one practical matter. Throughout these lectures, common
cosmological units will be employed in which $\hbar = c = k_b =
1$. All dimensionful quantities can then
be expressed as powers of an energy scale, commonly taken as GeV.  In
particular, length and time both have units of [GeV]$^{-1}$, while
Newton's constant $G$ has units of [GeV]$^{-2}$ since it is defined as
equal to the square of the inverse Planck mass.  These units are very
convenient for cosmology, because many problems deal with widely
varying scales simultaneously. For example, any computation of relic
particle abundances (e.g. primordial nucleosynthesis) involves both a
quantum mechanical scale (the interaction cross-section) and a
cosmological scale (the time scale for the expansion of the
universe). Conversion between these cosmological units and physical
(cgs) units can be achieved by inserting needed factors of $\hbar$, $c$,
and $k_b$.  The standard textbook by Kolb and Turner (1990) contains
an extremely useful appendix on units.

\subsection{Causes of temperature fluctuations}
\label{causes}

Blackbody radiation in a perfectly homogeneous and isotropic universe,
which is always adopted as a zeroth-order approximation, must be at a
uniform temperature, by assumption. When perturbations are introduced, three
elementary physical processes can produce a shift in the apparent
blackbody temperature of the radiation emitted from a particular point
in space. All temperature fluctuations in the microwave
background are due to one of the following three effects.

The first is simply a change in the intrinsic temperature of the
radiation at a given point in space. 
This will occur if the radiation density increases via
adiabatic compression, just as with the behavior of an ideal gas. The
fractional temperature perturbation in the radiation just equals the
fractional density perturbation.

The second is equally simple: a doppler shift if the radiation at
a particular point is moving with respect to the
observer. Any density
perturbations within the horizon scale will necessarily be accompanied
by velocity perturbations. The induced temperature perturbation in
the radiation equals the peculiar velocity (in units of $c$, of
course), with motion towards the observer corresponding to a positive
temperature perturbation.

The third is a bit more subtle: a difference in gravitational
potential between a particular point in space and an observer will
result in a temperature shift of the radiation propagating between the
point and the observer due to gravitational redshifting.
This is known as the Sachs-Wolfe effect, after the original
paper describing it (Sachs and Wolfe, 1967). This paper contains a
completely straightforward general relativistic calculation of the
effect, but the details are lengthy and complicated. A far simpler and
more intuitive derivation has been given by Hu and White (1997) making
use of gauge transformations. The Sachs-Wolfe effect is often broken
into two pieces, the usual effect and the so-called Integrated
Sachs-Wolfe effect. The latter
arises when gravitational potentials are evolving with time: radiation
propagates into a potential well, gaining energy and blueshifting in
the process. As it climbs out, it loses energy and redshifts, but if
the depth of the potential well has increased during the time the
radiation propagates through it, the redshift on exiting will be
larger than the blueshift on entering, and the radiation will gain a
net redshift, appearing cooler than it started out. Gravitational
potentials remain constant in time in a matter-dominated universe, so
to the extent the universe is matter dominated during the time the
microwave background radiation freely propagates, the Integrated
Sachs-Wolfe effect is zero. In models with significantly less than
critical density in matter (i.e. the currently popular $\Lambda$CDM
models), the redshift of matter-radiation equality occurs late enough
that the gravitational potentials are still evolving significantly
when the microwave background radiation decouples, leading to a
non-negligible Integrated Sachs-Wolfe effect. The same situation also
occurs at late times in these models; gravitational potentials begin
to evolve again as the universe makes a transition from matter
domination to either vacuum energy domination or
a significantly curved background spatial metric, 
giving an additional Integrated
Sachs-Wolfe contribution.

\subsection{A formal description}
\label{formalism}

The early universe at the epoch when the microwave background
radiation begins propagating freely, around a redshift of $z=1100$, is
a conceptually simple place. Its constituents are ``baryons'' (including
protons, helium nuclei, and electrons, even though electrons are not
baryons), neutrinos, photons, and dark
matter particles. The neutrinos
and dark matter can be treated as interacting only gravitationally
since their weak interaction cross-sections are too small at this
energy scale to be dynamically or thermodynamically relevant. The
photons and baryons interact electromagnetically, primarily via
Compton scattering of the radiation from the electrons. 
The typical interaction energies are low enough for the
scattering to be well-approximated by the simple Thomson cross
section. All other scattering processes (e.g. Thomson scattering from
protons, Rayleigh scattering of radiation from neutral hydrogen) have
small enough cross-sections to be insignificant, so we have four
species of matter with only one relevant (and simple) interaction
process among them.  The universe is also very close to being
homogeneous and isotropic, with small perturbations in density and
velocity on the order of a part in $10^5$. The tiny size of the
perturbations guarantees that linear perturbation theory 
around a homogeneous and isotropic
background universe will be an excellent approximation.

Conceptually, the formal description of the universe at this epoch is
quite simple. The unperturbed background cosmology is described by the
Friedmann-Robertson-Walker (FRW) metric, and the evolution of the
cosmological scale factor $a(t)$ in this metric is given by the
Friedmann equation (see the lectures by Peacock in this volume). The
evolution of the free electron density $n_e$ is determined by the
detailed atomic physics describing the recombination of neutral
hydrogen and helium; see Seager {\it et al.} (2000) for a detailed
discussion. At a temperature of around 0.5 eV, the electrons combine
with the protons and helium nuclei to make neutral atoms. As a result,
the photons cease Thomson scattering and propagate freely to us. The
microwave background is essentially an image of the ``surface of last
scattering''. Recombination must be calculated quite precisely because
the temperature and thickness of this surface depend sensitively on
the ionization history through the recombination process.

The evolution of first-order perturbations in the various energy
density components and the metric are described with the following
sets of equations:

\smallskip
\begin{itemize}

\item The photons and neutrinos are described by distribution
functions $f({\bf x}, {\bf p}, t)$. A fundamental simplifying
assumption is that the energy dependence of both is given by the
blackbody distribution. The space dependence is generally Fourier
transformed, so the distribution functions can be written as
$\Theta({\bf k}, {\bf\hat n}, t)$, where the function has been
normalized to the temperature of the blackbody distribution and ${\bf
\hat n}$ represents the direction in which the radiation
propagates. The time evolution of each is given by the Boltzmann
equation. For neutrinos, collisions are unimportant so the Boltzmann
collision term on the right side is zero; for photons, Thomson scattering off
electrons must be included. 

\item The dark matter and baryons are in principle described by
Boltzmann equations as well, but a fluid description incorporating
only the lowest two velocity moments of the distribution functions is
adequate. Thus each is described by the Euler and continuity equations
for their densities and velocities. The baryon Euler equation must
include the coupling to photons via Thomson scattering. 

\item Metric perturbation evolution and the connection of the metric
perturbations to the matter perturbations are both contained in the
Einstein equations.  This is where the subtleties arise. A general
metric perturbation has 10 degrees of freedom, but four of these are
unphysical gauge modes. The physical perturbations include two degrees
of freedom constructed from scalar functions, two from a vector, and
two remaining tensor perturbations (Mukhanov {\it et al.} 1992). Physically,
the scalar perturbations correspond to gravitational potential and
anisotropic stress perturbations; the vector perturbations correspond to 
vorticity and shear perturbations; and the tensor perturbations
are two polarizations of gravitational radiation. Tensor and vector
perturbations do not couple to matter evolving only under gravitation;
in the absence of a ``stiff source'' of stress energy, like cosmic
defects or magnetic fields, the tensor and vector perturbations
decouple from the linear perturbations in the matter.

\end{itemize}
 
A variety of different variable choices and methods for eliminating
the gauge freedom have been developed. The subject can be fairly
complicated. A detailed discussion and comparison between the
Newtonian and synchronous gauges, along with a complete set of
equations, can be found in Ma and Bertschinger (1995); also see Hu 
{\it et al.} (1998). An elegant and physically appealing formalism based on an
entirely covariant and gauge-invariant description of all physical
quantities has been developed for the microwave background by
Challinor and Lasenby (1999) and Gebbie {\it et al.} (2000), based on
earlier work by Ehlers (1993) and Ellis and Bruni (1989). A more conventional
gauge-invariant approach was originated by Bardeen (1980) and
developed by Kodama and Sasaki (1984). 

The Boltzmann equations are partial differential equations, which can
be converted to hierarchies of ordinary differential equations by
expanding their directional dependence in Legendre polynomials.  The
result is a large set of coupled, first-order linear ordinary
differential equations which form a well-posed initial value
problem. Initial conditions must be specified. Generally they are
taken to be so-called adiabatic perturbations: initial curvature
perturbations with equal fractional perturbations in each matter
species. Such perturbations arise naturally from the simplest
inflationary scenarios. Alternatively, isocurvature perturbations can
also be considered: these initial conditions have fractional density
perturbations in two or more matter species whose total spatial
curvature perturbation
cancels. The issue of numerically determining initial conditions is
discussed below in Sec.~\ref{initialconditions}. 

The set of equations are numerically stiff before last scattering, 
since they contain the two widely discrepant time scales: the
Thomson scattering time for electrons and photons, and the (much
longer) Hubble time. Initial conditions must be set with high accuracy
and an appropriate stiff integrator must be employed.
A variety of numerical techniques have been developed for evolving the
equations. Particularly important is the line-of-sight algorithm first
developed by Seljak and Zaldarriaga (1996) and then implemented by
them in the publicly-available CMBFAST code \hfil\break 
(see http://www.sns.ias.edu/\~{}matiasz/CMBFAST/cmbfast.html).

The above discussion is intentionally heuristic and somewhat vague
because many of the issues involved are technical and not particularly
illuminating. My main point is an appreciation for the detailed and
precise physics which goes into computing microwave background
fluctuations. However, all of this formalism should not obscure
several basic physical processes which determine the ultimate form of
the fluctuations. A widespread understanding of most of the physical
processes detailed below followed from a seminal paper by Hu and
Sugiyama (1994), a classic of the microwave background literature.

\subsection{Tight coupling}
\label{tightcoupling}

Two basic time scales enter into the evolution of the microwave
background. The first is the photon scattering time scale $t_s$, the
mean time between Thomson scatterings. The other is the expansion
time scale of the universe, $H^{-1}$, where $H={\dot a/a}$ is the
Hubble parameter. At temperatures significantly greater than 0.5 eV,
hydrogen and helium are completely ionized and $t_s \ll H^{-1}$. The
Thomson scatterings which couple the electrons and photons occur much
more rapidly than the expansion of the universe; as a result, the
baryons and photons behave as a single ``tightly coupled''
fluid. During this period, the fluctuations in the photons mirror the
fluctuations in the baryons. (Note that recombination occurs at around
0.5 eV rather than 13.6 eV because of the huge photon-baryon ratio;
the universe contains somewhere around $10^9$ photons for each baryon,
as we know from primordial nucleosynthesis. It is a useful exercise to
work out the approximate recombination temperature.)

The photon distribution function for scalar perturbations can be
written as $\Theta({\bf k}, \mu, t)$ where $\mu = {\bf\hat k}\cdot
{\bf\hat n}$ and the scalar character of the fluctuations guarantees
the distribution cannot have any azimuthal directional dependence.
(The azimuthal dependence for vector and tensor perturbations can also
be included in a similar decomposition).  The moments of the
distribution are defined as
\begin{equation}
\Theta({\bf k}, \mu, t) = \sum_{l=0}^\infty (-i)^l \Theta_l({\bf k},
t) P_l(\mu);
\label{moments}
\end{equation}
sometimes other normalizations are used. Tight coupling implies that
$\Theta_l = 0$ for $l > 1$. Physically, the $l=0$ moment corresponds
to the photon energy density perturbation, while $l=1$ corresponds to the bulk
velocity. During tight coupling, these two moments must match the
baryon density and velocity perturbations. Any higher moments rapidly
decay due to the isotropizing effect of Thomson scattering; this
follows immediately from the photon Boltzmann equation.

\subsection{Free streaming}
\label{freestreaming} 

In the other regime, for temperatures significantly lower than 0.5 eV,
$t_s \gg H^{-1}$ and photons on average never scatter again until the
present time. This is known as the ``free streaming'' epoch. Since the
radiation is no longer tightly coupled to the electrons, all higher
moments in the radiation field develop as the photons propagate. In a
flat background spacetime, the exact solution is simple to
derive. After scattering ceases, the photons evolve according to the
Liouville equation
\begin{equation}
\Theta' + ik\mu\Theta = 0
\label{liouville}
\end{equation}
with the trivial solution
\begin{equation}
\Theta({\bf k},\mu,\eta) = e^{-ik\mu(\eta - \eta_*)}\Theta({\bf k},\mu,\eta_*),
\label{thetafreestream}
\end{equation}
where we have converted to conformal time defined by $d\eta = dt/a(t)$
and $\eta_*$ corresponds to the time at which free streaming begins.
Taking moments of both sides results in
\begin{equation}
\Theta_l({\bf k},\eta) = (2l+1)\left[\Theta_0({\bf k},\eta_*)
j_l(k\eta - k\eta_*) + \Theta_1({\bf k},\eta_*) j_l'(k\eta -
k\eta_*)\right]
\label{momentfreestream}
\end{equation}
with $j_l$ a spherical Bessel function. The process of free streaming
essentially maps spatial variations in the photon distribution at the
last scattering surface (wavenumber $k$) into angular variations on
the sky today (moment $l$).

\subsection{Diffusion damping}
\label{diffusiondamping}

In the intermediate regime during recombination, $t_s\simeq
H^{-1}$. Photons propagate a characteristic distance $L_D$ during this
time. Since some scattering is still occurring, baryons experience a
drag from the photons as long as the ionization fraction is
appreciable. A second-order perturbation analysis shows that the
result is damping of baryon fluctuations on scales below $L_D$, known
as Silk damping or diffusion damping. This effect can be modelled
by the replacement
\begin{equation}
\Theta_0({\bf k},\eta_*) \rightarrow \Theta_0({\bf k},\eta_*) e^{-(kL_D)^2} 
\label{silkdamping}
\end{equation}
although detailed calculations are needed to define $L_D$
precisely. As a result of this damping, microwave background
fluctuations are exponentially suppressed on angular scales
significantly smaller than a degree.

\subsection{The resulting power spectrum}
\label{resultingpowerspectrum} 

The fluctuations in the universe are assumed to arise from
some random statistical process. We are not interested in the
exact pattern of fluctuations we see from our vantage point,
since this is only a single realization of the process. Rather,
a theory of cosmology predicts an underlying distribution, of
which our visible sky is a single statistical realization.
The most basic statistic describing fluctuations is their
power spectrum. A temperature map on the sky $T({\bf\hat n})$
is conventionally expanded in spherical harmonics,
\begin{equation}
     {T({\bf\hat n}) \over T_0}=1+\sum_{l=1}^\infty\sum_{m=-l}^l
     a^{\rm T}_{(lm)}\,Y_{(lm)}({\bf\hat n})
\label{Texpansion}
\end{equation}
where
\begin{equation}
 a^{\rm T}_{(lm)}={1\over T_0}\int d{\bf\hat n}\,T({\bf\hat n}) Y_{(lm)}^*({\bf\hat n})
\label{Tmoments}
\end{equation}
are the temperature multipole coefficients and $T_0$ is the
mean CMB temperature. 
The $l=1$ term in Eq.~(\ref{Texpansion}) is
indistinguishable from the kinematic dipole and is normally ignored.
The temperature angular power spectrum $C_l$ is then given by
\begin{equation}
\left\langle a^{{\rm T}\,*}_{(lm)}a^{\rm T}_{(l'm')}\right\rangle =
                C^{\rm T}_l \delta_{ll'}\delta_{mm'},
\label{Tpowerspectrum}
\end{equation}
where the angled brackets represent an average over statistical
realizations of the underlying distribution. Since we have only a
single sky to observe, an unbiased estimator of $C_l$ is constructed as
\begin{equation}
{\hat C}^{\rm T}_l = {1\over 2l+1}\sum_{m=-l}^l a^{{\rm T}\,*}_{lm} a^{\rm T}_{lm}.
\label{Clestimator}
\end{equation}
The statistical uncertainty in estimating $C^{\rm T}_l$ by a sum of $2l+1$
terms is known as ``cosmic variance''.
The constraints $l=l'$ and $m=m'$ follow from the
assumption of statistical isotropy: $C^{\rm T}_l$ must be independent of the
orientation of the coordinate system used for the harmonic
expansion. These conditions can be verified via an explicit rotation
of the coordinate system.

A given cosmological theory will predict $C^{\rm T}_l$ as a function
of $l$, which can be obtained from evolving the temperature
distribution function as described above. This prediction can then be
compared with data from measured temperature differences on the sky.
Figure 1 shows a typical temperature
power spectrum from the inflationary class of
models, described in more detail below. The distinctive sequence of
peaks arise from coherent acoustic oscillations in the fluid during
the tight coupling epoch and are of great importance in precision
tests of cosmological models; these peaks will be discussed in
Sec.~\ref{acousticoscillations}.
The effect of diffusion damping is clearly visible in the decreasing
power above $l=1000$. When viewing angular power spectrum plots in
multipole space, keep in mind that $l=200$ corresponds approximately
to fluctuations on angular scales of a degree, and the angular scale
is inversely proportional to $l$. The vertical axis is conventionally
plotted as $l(l+1)C^{\rm T}_l$ because the Sachs-Wolfe temperature
fluctuations from a scale-invariant spectrum of density perturbations
appears as a horizontal line on such a plot.

\begin{figure}
\plotone{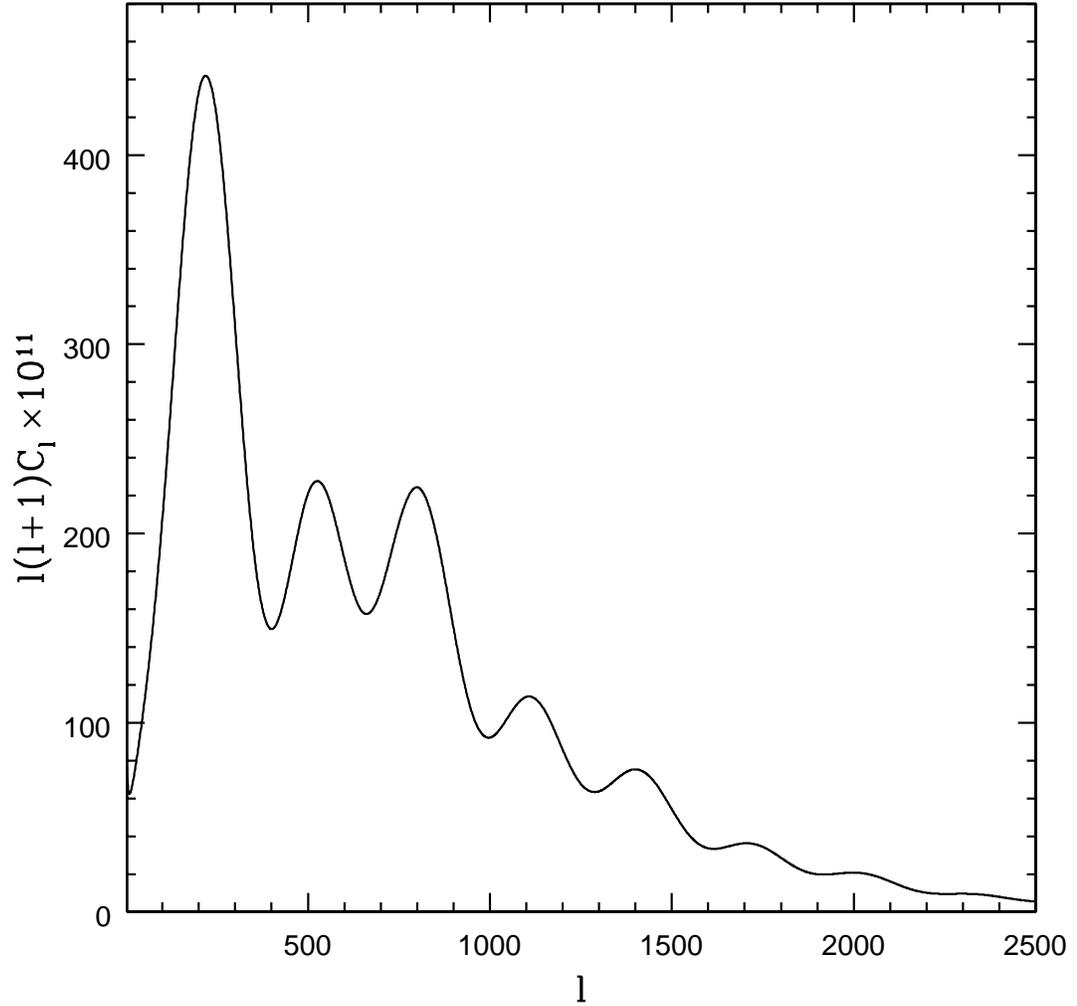}
\caption{The temperature angular power spectrum for a cosmological
model with mass density $\Omega_0 = 0.3$, vacuum
energy density $\Omega_\Lambda = 0.7$, 
Hubble parameter $h=0.7$, and a scale-invariant spectrum
of primordial adiabatic perturbations. \label{fig1}}
\end{figure}

\section{Physics of Polarization Fluctuations}
\label{polarizationsection}

In addition to temperature fluctuations, the simple physics of 
decoupling inevitably leads to non-zero polarization of the microwave
background radiation as well, although quite generically the
polarization fluctuations are expected to be significantly smaller
than the temperature fluctuations. This section reviews the physics of
polarization generation and its description. For a more detailed
pedagogical discussion of microwave background polarization, see
Kosowsky (1999), from which this section is excerpted.

\subsection{Stokes parameters}
\label{stokesparameters}

Polarized light is conventionally described in terms of the
Stokes parameters, which are presented in any optics text.
If a monochromatic electromagnetic wave
propogating in
the $z$-direction has an electric field vector at a given
point in space given by 
\begin{equation}
E_x=a_x(t)\cos\left[\omega_0 t - \theta_x(t)\right],\quad
E_y=a_y(t)\cos\left[\omega_0 t - \theta_y(t)\right],
\label{efield}
\end{equation}
then the Stokes parameters are defined as the following time
averages: 
%\begin{mathletters}
\label{stokesdef}
\begin{eqnarray}
I\,&&\equiv \langle a_x^2\rangle + \langle a_y^2\rangle;\\
Q\,&&\equiv \langle a_x^2\rangle - \langle a_y^2\rangle;\\
U\,&&\equiv \langle 2a_xa_y\cos(\theta_x -\theta_y)\rangle;\\
V\,&&\equiv \langle 2a_xa_y\sin(\theta_x -\theta_y)\rangle.
\end{eqnarray}
%\end{mathletters}
The averages are over times long compared to the inverse
frequency of the wave.
The parameter $I$ gives the intensity of the radiation which is
always positive and is equivalent to the temperature for
blackbody radiation. The other three parameters define the
polarization state of the wave and can have either sign. 
Unpolarized radiation, or ``natural light,'' is described by
$Q=U=V=0$. 

The parameters $I$ and $V$ are physical observables independent
of the coordinate system, but $Q$ and $U$ depend on the
orientation of the $x$ and $y$ axes. If a given wave is
described by the parameters $Q$ and $U$ for a certain
orientation of the coordinate system, then after a rotation of
the $x-y$ plane through an angle $\phi$, it is
straightforward to verify that the same wave is now
described by the parameters
\begin{eqnarray}
Q'&&=Q\cos(2\phi) + U\sin(2\phi),\nonumber\\
U'&&=-Q\sin(2\phi) + U\cos(2\phi).
\label{uqtransf}
\end{eqnarray}
{}From this transformation it is easy to see that the quantity
$P^2 \equiv Q^2+U^2$ is invariant under rotation of the axes, and the angle
\begin{equation}
\alpha\equiv{1\over 2}\tan^{-1}{U\over Q}
\label{alpha}
\end{equation}
defines a constant orientation parallel
to the electric field of the wave.
The Stokes parameters are a useful description of
polarization because they are {\it additive} for incoherent
superposition of radiation; note this is not true for the
magnitude or orientation of polarization.
Note that the transformation law in Eq.~(\ref{uqtransf}) is characteristic
not of a vector but of the second-rank {\it tensor}
\begin{equation}
\rho={1\over 2} \pmatrix{\vphantom\int I+Q & U-iV\cr 
\vphantom\int U+iV& I-Q},
\label{densmatrix}
\end{equation}
which also corresponds to the quantum mechanical
density matrix for an ensemble of photons (Kosowsky 1996).
In kinetic theory, the photon distribution function
$f({\bf x}, {\bf p}, t)$ discussed in Sec.~\ref{formalism} must
be generalized to $\rho_{ij}({\bf x}, {\bf p}, t)$, corresponding
to this density matrix.

\subsection{Thomson scattering and the quadrupolar source}
\label{thomsonscattering}

Non-zero linear polarization in the microwave background is generated around
decoupling because the Thomson scattering which couples the radiation
and the electrons is not isotropic but varies with the scattering
angle. The total scattering
cross-section, defined as the radiated intensity per unit solid
angle divided by the incoming intensity per unit area, is given by
\begin{equation}
{d\sigma\over d\Omega} = {3\sigma_T\over 8\pi} 
\left|{\hat\varepsilon}'\cdot{\hat\varepsilon}\right|^2
\label{thomson}
\end{equation}
where $\sigma_T$ is the total Thomson cross section and
the vectors $\hat\varepsilon$ and ${\hat\varepsilon}'$ are unit
vectors in the planes perpendicular to the propogation directions
which are aligned with the outgoing and incoming polarization,
respectively. This scattering cross-section can give no net
circular polarization, so $V=0$ for cosmological perturbations and
will not be discussed further. Measurements of $V$ polarization can
be used as a diagnostic of systematic errors or microwave foreground
emission.

It is a straightforward but slightly involved exercise to show that
the above relations imply that an incoming unpolarized radiation field
with the multipole expansion Eq.~(\ref{Texpansion}) incident
upon an electron in a sample volume with cross-section $\sigma_B$ 
will be Thomson scattered into an outgoing radiation field 
from the sample volume with Stokes parameters
\begin{equation}
Q({\bf\hat n})-iU({\bf\hat n}) 
={3\sigma_T\over 8\pi\sigma_B} \sqrt{\pi\over 5}\,a_{20}\sin^2\beta
\label{a20}
\end{equation}
if the incoming radiation propagating in a given direction 
has rotational symmetry around its
direction of propagation, as will hold for individual Fourier modes of
scalar perturbations. Explicit expressions for the general case of no
symmetry can be derived in terms of Wigner D-symbols (Kosowsky 1999). 

In simple and general terms, unpolarized incoming radiation will be
Thomson scattered into linearly polarized radiation if and only if the
incoming radiation  has a non-zero quadrupolar directional dependence.
This single fact is sufficient to understand the fundamental physics
behind polarization of the microwave background. During the
tight-coupling epoch, the radiation field has only monopole and dipole
directional dependences as explained above; therefore, scattering can
produce no net polarization and the radiation remains unpolarized.
As tight coupling begins to break down as recombination begins, 
a quadrupole moment of the
radiation field will begin to grow due to free streaming of the photons.
Polarization is generated during the brief interval when a significant
quadrupole moment of the radiation has built up, but the scattering
electrons have not yet all recombined. Note that if the universe
recombined instantaneously, the net polarization of the microwave
background would be zero. Due to this competition between the
quadrupole source building up and the density of scatterers declining,
the amplitude of polarization in the microwave background is
generically suppressed by an order of magnitude compared to the
temperature fluctuations.

Before polarization generation commences, the temperature fluctuations
have either a monopole dependence, corresponding to density
perturbations, or a dipole dependence, corresponding to velocity
perturbations. A straightforward solution to the photon free-streaming
equation (in terms of spherical Bessel functions) shows that for
Fourier modes with wavelengths large compared to a characteristic
thickness of the surface of last scattering, the quadrupole
contribution through the last scattering surface is dominated by the
velocity fluctuations in the temperature, not the density
fluctuations. This makes intuitive sense: the dipole fluctuations can
free stream directly into the quadrupole, but the monopole
fluctuations must stream through the dipole first. This conclusion
breaks down on small scales where either monopole or dipole
can be the dominant quadrupole source, but numerical computations show
that on scales of interest for microwave background fluctuations, 
the dipole temperature fluctuations are always the
dominant source of quadrupole fluctuations at the surface of last
scattering. Therefore, polarization fluctuations reflect mainly
velocity perturbations at last scattering, in contrast to temperature
fluctuations which predominantly reflect density perturbations.

\subsection{Harmonic expansions and power spectra}
\label{harmonicexpansions}

Just as the temperature on the sky can be expanded into spherical
harmonics, facilitating the computation of the angular power spectrum,
so can the polarization. The situation is formally parallel, although
in practice it is more complicated: while the temperature is a scalar
quantity, the polarization is a second-rank tensor. We can define
a polarization tensor with the correct transformation properties,
Eq.~(\ref{uqtransf}), as
\begin{equation}
  P_{ab}({\bf\hat n})={1\over 2} \left( \begin{array}{cc}
   Q({\bf\hat n}) & -U({\bf\hat n}) \sin\theta \\
   \noalign{\vskip6pt}
   - U({\bf\hat n})\sin\theta & -Q({\bf\hat n})\sin^2\theta \\
   \end{array} \right).
\label{whatPis}
\end{equation}
The dependence on the Stokes parameters is the same as for the density
matrix, Eq.~\ref{densmatrix}; the extra factors are convenient because
the usual spherical coordinate basis is orthogonal but not
orthonormal. This tensor quantity must be expanded in terms of tensor
spherical harmonics which preserve the correct transformation
properties. We assume a complete set of orthonormal basis functions
for symmetric trace-free $2\times2$ tensors on the sky,
\begin{equation}
     {P_{ab}({\bf\hat n})\over T_0} = \sum_{l=2}^\infty\sum_{m=-l}^l
     \left[ a_{(lm)}^{\rm G}
     Y_{(lm)ab}^{\rm G}({\bf\hat n}) + a_{(lm)}^{\rm C} Y_{(lm)ab}^{\rm C}
     ({\bf\hat n}) \right],
\label{Pexpansion}
\end{equation}
where the expansion coefficients are given by
\begin{eqnarray}
     a_{(lm)}^{\rm G}&=&{1\over T_0}\int \, d{\bf\hat n} P_{ab}({\bf\hat n})
               Y_{(lm)}^{{\rm G} \,ab\, *}({\bf\hat n}), \\
     a_{(lm)}^{\rm C}&=&{1\over T_0}\int d{\bf\hat n}\, P_{ab}({\bf\hat n})
                             Y_{(lm)}^{{\rm C} \, ab\, *}({\bf\hat n}),
\label{defmoments}
\end{eqnarray}
which follow from the orthonormality properties
\begin{equation}
 \int d{\bf\hat n}\,Y_{(lm)ab}^{{\rm G}\,*}({\bf\hat n})\,
       Y_{(l'm')}^{{\rm G}\,\,ab}({\bf\hat n})
=\int d{\bf\hat n}\,Y_{(lm)ab}^{{\rm C}\,*}({\bf\hat n})\,
       Y_{(l'm')}^{{\rm C}\,\,ab}({\bf\hat n})
=\delta_{ll'} \delta_{mm'},\nonumber
\end{equation}
\begin{equation}
\int d{\bf\hat n}\,Y_{(lm)ab}^{{\rm G}\, *}({\bf\hat n})\,
Y_{(l'm')}^{{\rm C}\,\,ab}({\bf\hat n})
=0.
\label{norms}
\end{equation}

These tensor spherical harmonics are not as exotic as they might
sound; they are used extensively in the theory of gravitational
radiation, where they naturally describe the radiation multipole
expansion.  Tensor spherical harmonics are similar to vector spherical
harmonics used to represent electromagnetic radiation fields, familiar
from Chapter 16 of Jackson (1975).  Explicit formulas for tensor
spherical harmonics can be derived via various algebraic and group
theoretic methods; see Thorne (1980) for a complete discussion.  A
particularly elegant and useful derivation of the tensor spherical
harmonics (along with the vector spherical harmonics as well) is
provided by differential geometry: the harmonics can be
expressed as covariant derivatives of the usual spherical harmonics
with respect to an underlying manifold of a two-sphere (i.e., the
sky).  This construction has been carried out explicitly and applied
to the microwave background polarization (Kamionkowski, Kosowsky, and
Stebbins 1996).

The existence of two sets of basis functions, labeled here by ``G''
and ``C'', is due to the fact that the symmetric traceless $2\times2$
tensor describing linear polarization is specified by two independent
parameters.  In two dimensions, any symmetric traceless tensor can be
uniquely decomposed into a part of the form $A_{;ab}-(1/2)g_{ab}
A_{;c}{}^c$ and another part of the form
$B_{;ac}\epsilon^c{}_b+B_{;bc}\epsilon^c{}_a$ where $A$ and $B$ are
two scalar functions and semicolons indicate covariant derivatives.  This
decomposition is quite similar to the decomposition of a vector field
into a part which is the gradient of a scalar field and a part which
is the curl of a vector field; hence we use the notation G for
``gradient'' and C for ``curl''. In fact, this correspondence is more
than just cosmetic: if a linear polarization field is visualized in
the usual way with headless ``vectors'' representing the amplitude and
orientation of the polarization, then the G harmonics describe the
portion of the polarization field which has no handedness associated
with it, while the C harmonics describe the other portion of the field
which does have a handedness (just as with the gradient and curl of a
vector field). Note that Zaldarriaga and Seljak (1997) label these
harmonics E and B, with a slightly different normalization than
defined here (see Kamionkowski {\it et al.}\ 1996). 

We now have three sets of multipole moments, $a_{(lm)}^{\rm T}$,
$a_{(lm)}^{\rm G}$, and $a_{(lm)}^{\rm C}$,
which fully describe the
temperature/polarization map of the sky. These moments can be combined
quadratically into various power spectra analogous to the temperature
$C_l^{\rm T}$. 
Statistical isotropy implies that
\begin{eqnarray}
\left\langle a_{(lm)}^{{\rm T}\,*}a_{(l'm')}^{\rm T}\right\rangle =
                C^{\rm T}_l \delta_{ll'}\delta_{mm'},&\qquad\qquad&
\left\langle a_{(lm)}^{{\rm G}\,*}a_{(l'm')}^{\rm G}\right\rangle = 
                                C_l^{\rm G} \delta_{ll'}\delta_{mm'},
                                    \cr
\left\langle a_{(lm)}^{{\rm C}\,*}a_{(l'm')}^{\rm C}\right\rangle =
                C_l^{\rm C} \delta_{ll'}\delta_{mm'},&\qquad\qquad&
\left\langle a_{(lm)}^{{\rm T}\,*}a_{(l'm')}^{\rm G}\right\rangle =
                               C_l^{\rm TG}\delta_{ll'}\delta_{mm'},
                                      \cr
\left\langle a_{(lm)}^{{\rm T}\,*}a_{(l'm')}^{\rm C}\right\rangle =
                C_l^{\rm TC}\delta_{ll'}\delta_{mm'},&\qquad\qquad&
\left\langle a_{(lm)}^{{\rm G}\,*}a_{(l'm')}^{\rm C}\right\rangle =
                               C_l^{\rm GC}\delta_{ll'}\delta_{mm'},
\label{cldefs}
\end{eqnarray}
where the angle brackets are an average over all realizations of the
probability distribution for the cosmological initial conditions.
Simple statistical estimators of the various $C_l$'s can be
constructed from maps of the microwave background temperature
and polarization.

For fluctuations with gaussian random distributions
(as predicted by the simplest inflation models), 
the statistical properties of a
temperature/polarization map are specified fully by these six
sets of multipole moments.
In addition, the scalar spherical harmonics $Y_{(lm)}$ and the G
tensor harmonics $Y_{(lm)ab}^{\rm G}$ have parity $(-1)^l$, but the C
harmonics $Y_{(lm)ab}^{\rm C}$ have parity $(-1)^{l+1}$.
If the large-scale perturbations in the early universe were invariant
under parity inversion, then
$C_l^{\rm TC}=C_l^{\rm GC}=0$. So generally, microwave background
fluctuations are characterized by the four power spectra
$C_l^{\rm T}$, $C_l^{\rm G}$, $C_l^{\rm C}$, and $C_l^{\rm TG}$. These
The end result of the numerical computations described
in Sec.~\ref{formalism} above are these power spectra. 
Polarization power spectra $C_l^{\rm G}$ and $C_l^{\rm TG}$ for scalar
perturbations in a typical inflation-like cosmological
model, generated with the
CMBFAST code (Seljak and Zaldarriaga 1996),
are displayed in Fig.~2.
The temperature power spectrum in Fig.~1 and
the polarization power spectra in Fig.~2 come from the
same cosmological model. The physical source of the 
features in the power spectra
is discussed in the next section, followed by a discussion of
how cosmological parameters can be determined to high precision 
via detailed measurements of the microwave background power spectra. 

\begin{figure}
\plotone{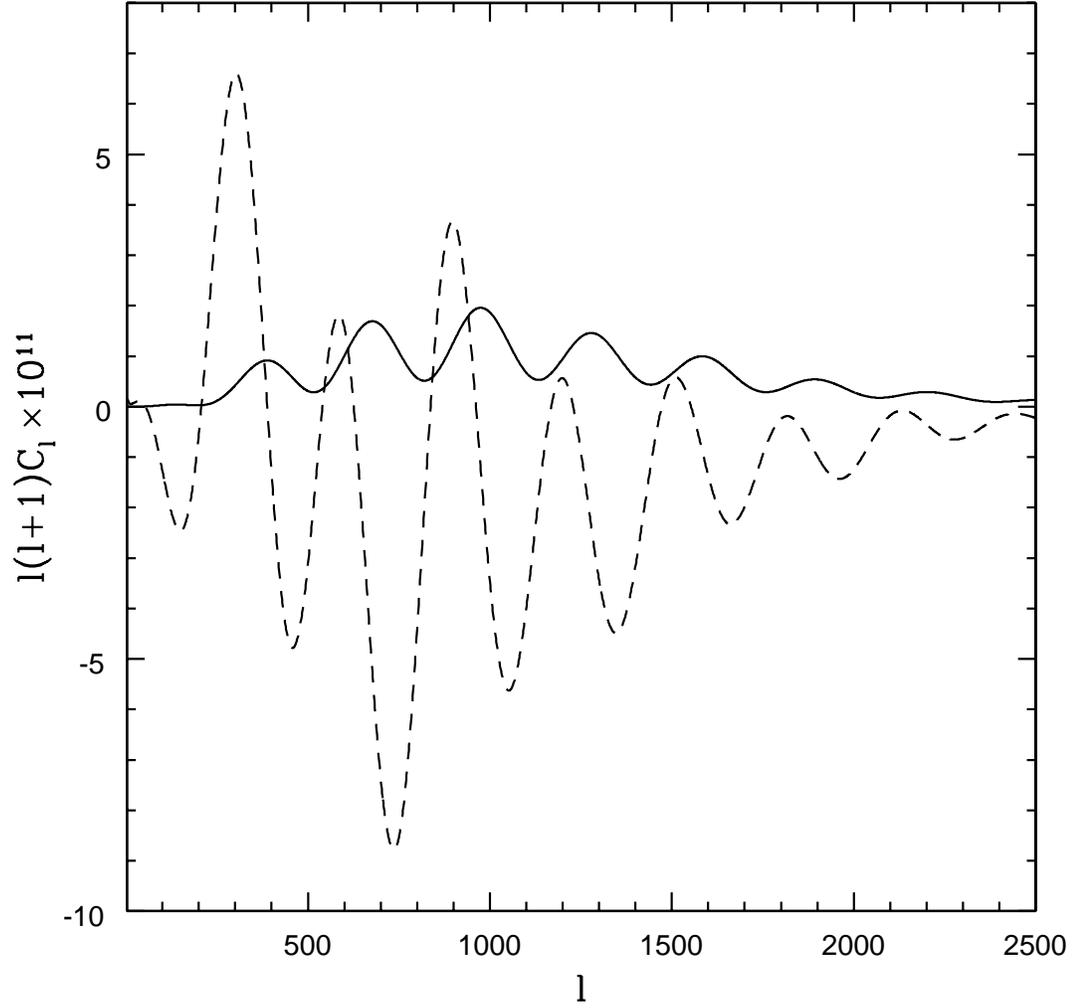}
\caption{The G polarization power spectrum (solid line) and the cross-power
TG between temperature and polarization (dashed line), for the same model as 
in Fig.~1. \label{fig2}}
\end{figure}

\section{Acoustic Oscillations}
\label{acousticoscillations}

Before decoupling, the matter in the universe has significant
pressure because it is tightly coupled to radiation. This pressure
counteracts any tendency for matter to collapse
gravitationally. Formally, the Jeans mass is greater than the mass
within a horizon volume for times earlier than decoupling. During this
epoch, density perturbations will set up standing acoustic waves in
the plasma. Under certain conditions, these waves leave a distinctive
imprint on the power spectrum of the microwave background, which in
turn provides the basis for precision constraints on cosmological
parameters. This section reviews the basics of the acoustic
oscillations. 

\subsection{An oscillator equation}
\label{oscillatorequation}

In their classic 1996 paper, Hu and Sugiyama transformed the basic
equations describing the evolution of perturbations into an oscillator
equation. Combining the zeroth moment of the photon Boltzmann equation
with the baryon Euler equation for a given $k$-mode
in the tight-coupling approximation
(mean baryon velocity equals mean radiation velocity) gives
\begin{equation}
{\ddot\Theta_0} + H{R\over 1+R}{\dot\Theta_0} + k^2 c_s^2\Theta_0
= -\ddot\Phi - H{R\over 1+R}\dot\Phi - {1\over 3}k^2\Psi,
\label{oscillator}
\end{equation}
where $\Theta_0$ is the zeroth moment of the temperature distribution
function (proportional to the photon density perturbation), 
$R=3\rho_b/4\rho_\gamma$ is proportional to the scale factor $a$, $H =
{\dot a}/a$ is the conformal Hubble parameter, and the sound speed is
given by $c_s^2 = 1/(3+3R)$. (All overdots are derivatives with
respect to conformal time.) $\Phi$ and $\Psi$ are the scalar metric 
perturbations in the Newtonian gauge; if we neglect the anisotropic
stress, which is generally small in conventional cosmological
scenarios, then $\Psi = -\Phi$. But the details are not very
important. The equation represents damped, driven oscillations of
the radiation density, and the various physical effects are
easily identified. The second term on the left side is the damping
of oscillations due to the expansion of the universe. The third
term on the left side is the restoring force due to the pressure,
since $c_s^2 = dP/d\rho$. On the right side, the first two terms
depend on the time variation of the gravitational potentials, so these
two are the source of the Integrated Sachs-Wolfe effect. The final
term on the right side is the driving term due to the gravitational
potential perturbations. As Hu and Sugiyama emphasized, these damped,
driven acoustic oscillations account for all of the structure in the
microwave background power spectrum.

A WKB approximation to the homogeneous equation with no driving source
terms gives the two oscillation modes (Hu and Sugiyama 1996)
\begin{equation}
\Theta_0(k,\eta) \propto \cases{(1+R)^{-1/4}\cos kr_s(\eta)\cr
                                (1+R)^{-1/4}\sin kr_s(\eta)}
\label{wkbmodes}
\end{equation}
where the sound horizon $r_s$ is given by
\begin{equation}
r_s(\eta) \equiv \int_0^\eta c_s(\eta') d\eta'.
\label{soundhorizon}
\end{equation}
Note that at times well before matter-radiation equality, the sound speed is
essentially constant, $c_s=1/\sqrt{3}$, and the sound horizon is
simply proportional to the causal horizon. 
In general, any perturbation with wavenumber $k$ will set up an
oscillatory behavior in the primordial plasma described by a linear
combination of the two modes in Eq.~(\ref{wkbmodes}). The relative
contribution of the modes will be determined by the initial conditions
describing the perturbation.

Equation (\ref{oscillator}) appears to be simpler than it actually
is, because $\Phi$ and $\Psi$ are the total gravitational potentials
due to all matter and radiation, including the photons which the
left side is describing. In other words, the right side of the
equation contains an implicit dependence on $\Theta_0$. At the expense
of pedagogical transparency, this situation can be remedied by
considering separately the potential from the photon-baryon fluid and
the potential from the truly external sources, the dark matter and
neutrinos. This split has been performed by Hu and White (1996). The
resulting equation, while still an oscillator equation, is much more
complicated, but must be used for a careful physical analysis of
acoustic oscillations.

\subsection{Initial conditions}
\label{initialconditions}

The initial conditions for radiation perturbations for a given
wavenumber $k$ can be broken into two categories, according to whether
the gravitational potential perturbation from the baryon-photon fluid,
$\Phi_{b\gamma}$, is nonzero or zero as $\eta\rightarrow 0$. The
former case is known as ``adiabatic'' (which is somewhat of a misnomer
since adiabatic technically refers to a property of a time-dependent process)
and implies that $n_b/n_\gamma$, the ratio of baryon to photon number
densities, is a constant in space. This case must couple to the cosine
oscillation mode since it requires $\Theta_0\neq 0$ as
$\eta\rightarrow 0$. The simplest (i.e. single-field) models of
inflation produce perturbations with adiabatic initial conditions.

The other case is termed ``isocurvature'' since the fluid
gravitational potential perturbation $\Phi_{b\gamma}$, and hence the
perturbations to the spatial curvature, are zero. 
In order to arrange such a perturbation,
the baryon and photon densities must vary in such a way that they
compensate each other: $n_b/n_\gamma$ varies, and thus these
perturbations are in entropy, not curvature. At an early enough time,
the temperature perturbation in a given $k$ mode must arise entirely
from the Sachs-Wolfe effect, and thus isocurvature perturbations
couple to the sine oscillation mode. These perturbations arise from
causal processes like phase transitions: a phase transition cannot
change the energy density of the universe from point to point, but it
can alter the relative entropy between various types of matter
depending on the values of the fields involved.
The potentially most interesting cause of isocurvature perturbations
is multiple dynamical fields in inflation. The fields will exchange
energy during inflation, and the field values will vary stochastically
between different points in space at the end of the phase transition,
generically giving isocurvature along with adiabatic perturbations
(Polarski and Starobinsky 1994).

The numerical problem of setting initial conditions is somewhat
tricky. The general problem of evolving perturbations involves linear
evolution equations for around a dozen variables, outlined in
Sec.~\ref{formalism}. Setting the correct
initial conditions involves specifying the value of each variable in
the limit as $\eta \rightarrow 0$. This is difficult for two reasons:
the equations are singular in this limit, and the equations become
increasingly numerically stiff in this limit. Simply using the
leading-order asymptotic behavior for all of the variables is only
valid in the high-temperature limit. Since the equations are stiff, small
departures from this limiting behavior 
in any of the variables can lead to numerical instability until
the equations evolve to a stiff solution, and this numerical solution
does not necessarily correspond to the desired initial
conditions. Numerical techniques for setting the initial conditions to high
accuracy at temperaturesare currently being developed.

\subsection{Coherent oscillations}
\label{coherentoscillations}

The characteristic ``acoustic peaks'' which appear in Figure 1 arise
from acoustic oscillations which are phase coherent: 
at some point in time, the phases of all of the acoustic oscillations
were the same. This requires the same initial
condition for {\it all} $k$-modes, including those with wavelengths
longer than the horizon. Such a condition arises naturally for
inflationary models, but is very hard to reproduce in models producing
perturbations causally on scales smaller than the horizon. Defect
models, for example, produce acoustic oscillations, but the
oscillations generically have incoherent phases and thus display no
peak structure in their power spectrum (Seljak {\it et al.} 1997). 
Simple models of inflation which produce only adiabatic perturbations
insure that all perturbations have the same phase at $\eta=0$ because
all of the perturbations are in the cosine mode of Eq.~(\ref{wkbmodes}).

A glance at the $k$ dependence of the adiabatic perturbation mode
reveals how the coherent peaks are produced. The microwave background
images the radiation density at a fixed time; as a function of $k$,
the density varies like $\cos(kr_s)$, where $r_s$ is
fixed. Physically, on scales much larger than the horizon at
decoupling, a perturbation mode has not had enough time to evolve.
At a particular smaller scale, the perturbation mode evolves to its
maximum density in potential wells, at which point decoupling occurs.
This is the scale reflected in the first acoustic peak in the power
spectrum. Likewise, at a particular still smaller scale, the
perturbation mode evolves to its maximum density in potential wells and
then turns around, evolving to its minimum density in potential wells;
at that point, decoupling occurs. This scale corresponds to that of
the second acoustic peak. (Since the power spectrum is the square of
the temperature fluctuation, both compressions and rarefactions in
potential wells correspond to peaks in the power spectrum.) 
Each successive peak represents successive
oscillations, with the scales of odd-numbered peaks corresponding to
those perturbation scales which have ended up compressed in potential
wells at the time of decoupling, while the even-numbered peaks
correspond to the perturbation scales which are rarefied in potential
wells at decoupling. If the perturbations are not phase coherent, then
the phase of a given $k$-mode at decoupling is not well defined, and
the power spectrum just reflects some mean fluctuation power at that
scale. 

In practice, two additional effects must be
considered: a given scale in $k$-space is mapped to a range of
$l$-values; and radiation velocities as well as densities contribute
to the power spectrum. The first effect broadens out the peaks, while
the second fills in the valleys between the peaks since the velocity
extrema will be exactly out of phase with the density extrema.
The amplitudes of the peaks in the power spectrum are also suppressed by
Silk damping, as mentioned in Sec.~\ref{diffusiondamping}. 

\subsection{The effect of baryons}
\label{baryoneffect}

The mass of the baryons creates a distinctive signature in the
acoustic oscillations (Hu and Sugiyama 1996). The zero-point of the oscillations is obtained
by setting $\Theta_0$ constant in Eq.~(\ref{oscillator}): the result is
\begin{equation}
\Theta_0 \simeq {1\over 3c_s^2}\Phi = (1+a)\Phi.
\label{zeropoint}
\end{equation}
The photon temperature $\Theta_0$ is not itself observable, but must
be combined with the gravitational redshift to form the ``apparent
temperature'' $\Theta_0 - \Phi$, which oscillates around $a\Phi$. 
If the oscillation amplitude is much larger than $a\Phi =
3\rho_b\Phi/4\rho_\gamma$, then the oscillations are effectively about
the mean temperature. The positive and negative oscillations are of
the same amplitude, so when the apparent temperature is squared to form the
power spectrum, all of the peaks have the same height. On the other
hand, if the baryons contribute a significant mass so that $a\Phi$ is 
a significant fraction of the oscillation amplitude, then the zero
point of the oscillations are displaced, and when the apparent
temperature is squared to form the power spectrum, the peaks arising
from the positive oscillations are higher than the peaks from the
negative oscillations. If $a\Phi$ is larger than the amplitude of the
oscillations, then the power spectrum peaks corresponding to the
negative oscillations disappear entirely. The physical interpretation
of this effect is that the baryon mass deepens the potential well in
which the baryons are oscillating, increasing the compression of the
plasma compared to the case with less baryon mass. In short, as the baryon
density increases, the power spectrum peaks corresponding to
compressions in potential wells get higher, while the alternating
peaks corresponding to rarefactions get lower. This alternating peak
height signature is a distinctive signature of baryon mass, and allows
the precise determination of the cosmological baryon density with the 
measurement of the first several acoustic peak heights.

\section{Cosmological Models and Constraints}
\label{cosmologicalmodels}

The cosmological interpretation of a measured microwave background
power spectrum requires, to some extent, 
the introduction of a particular space of models. A very simple,
broad, and well-motivated set of models are motivated by inflation:
a universe described by a homogeneous and isotropic background with
phase-coherent, power-law initial perturbations which evolve freely.
This model space excludes, for example, perturbations caused by
topological defects or other ``stiff'' sources, arbitrary initial power
spectra, or any departures from the standard background cosmology.
This set of models has the twin virtues of being relatively simple to 
calculate and best conforming to current power spectrum measurements. 
(In fact, most competing cosmological models, like those employing
cosmic defects to make structure, are essentially ruled out by current
microwave background and large-scale structure measurements.) This
section will describe the parameters defining the model space and
discuss the extent to which the parameters can be constrained through
the microwave background. 

\subsection{A space of models}
\label{spaceofmodels}

The parameters defining the model space can be broken into three
types: cosmological parameters describing the background space-time;
parameters describing the initial conditions; and other parameters
describing miscellaneous additional physical effects.
Background cosmological parameters are:

\begin{itemize}

\item $\Omega$, the ratio of the total energy density to the critical
density $\rho_c = 8\pi/3H^2$.  This parameter determines the spatial
curvature of the universe: $\Omega=1$ is a flat universe with critical
density. Smaller values of $\Omega$ correspond to a negative spatial
curvature, while larger values correspond to positive curvature. 
Current microwave background measurements constrain $\Omega$ to be
roughly within the rage 0.8 to 1.2, consistent with a critical-density
universe.  

\item $\Omega_b$, the ratio of the baryon density to the
critical density. Observations of the abundance of deuterium in high
redshift gas clouds and comparison with predictions from primordial
nucleosynthesis place strong constraints on this parameter (Tytler 
{\it et al.} 2000).  

\item $\Omega_m$, the ratio of the dark matter density to the critical
density. Dynamical constraints, gravitational lensing, cluster
abundances, and numerous other lines of evidence all point to a total
matter density in the neighborhood of 
$\Omega_0 = \Omega_m + \Omega_b = 0.3$. 

\item $\Omega_\Lambda$, the ratio of vacuum energy density $\Lambda$ 
to the critical
density. This is the notorious cosmological constant. Several years
ago, almost no cosmologist advocated a cosmological constant; now
almost every cosmologist accepts its existence. The shift was
precipitated by the Type Ia supernova Hubble diagram (Perlmutter {\it et al.} 
1999, Riess {\it et al.} 1998) which shows an apparent acceleration in the
expansion of the universe. Combined with strong constraints on
$\Omega$, a cosmological constant now seems unavoidable, although
high-energy theorists have a difficult time accepting it. Strong
gravitational lensing of quasars places upper limits on $\Omega_\Lambda$
(Falco {\it et al.} 1998). 

\item The present Hubble parameter $h$, in units of 100
km/s/Mpc. Distance ladder measurements (Mould {\it et al.} 2000) and
supernova Ia measurements (Riess {\it et al.} 1998)
give consistent estimates for $h$ of around
0.70, with systematic errors on the order of 10\%. 

\item Optionally, further parameters describing additional
contributions to the energy density of the universe. An example is
``quintessence'' models (Caldwell {\it et al.} 1998) which add one or more
scalar fields to the universe.  

\end{itemize}

\noindent
Parameters describing the initial conditions are:

\begin{itemize}

\item The amplitude of fluctuations $Q$, often defined at the
quadrupole scale. COBE fixed this amplitude to high accuracy
(Bennett {\it et al.} 1996). 

\item The power law index $n$ of initial adiabatic density fluctuations.
The scale-invariant Harrison-Zeldovich spectrum is $n=1$. Comparison
of microwave background and large-scale structure measurements shows
that $n$ is close to unity.

\item The relative contribution of tensor and scalar perturbations
$r$, usually defined as the ratio of the power at $l=2$ from each type
of perturbation. The fact that prominent features are seen in the
power spectrum (presumably arising from scalar density perturbations)
limits the power spectrum contribution of tensor perturbations to
roughly 20\% of the scalar amplitude. 

\item The power law index $n_T$ of tensor
perturbations. Unfortunately, tensor power spectra are generally
defined so that $n_T=0$ corresponds to scale invariant, in contrast to
the scalar case. 

\item Optionally, more parameters describing either departures of the
scalar perturbations from a power law (e.g. Kosowsky and Turner 1995)
or a small admixture of isocurvature perturbations.

\end{itemize}

\noindent
Other miscellaneous parameters include:

\begin{itemize}

\item A significant neutrino mass $m_\nu$. None of the current
neutrino oscillation results favor a cosmologically interesting
neutrino mass.  

\item The effective number of neutrino species $N_\nu$. This quantity
includes any particle species which is relativistic when it decouples
or can model entropy production prior to last scattering.

\item The redshift of reionization, $z_r$. Spectra of quasars at
redshift $z=5$ show that the universe has been reionized at least
since then.  

\end{itemize}

A realistic parameter analysis might include at least 8 free
parameters. Given a particular microwave
background measurement, deciding on a particular set of parameters and
various priors on those parameters is as much art as science. For the
correct model, parameter values should be insensitive to the size of
the parameter space or the particular priors invoked. Several
particular parameter space analyses are mentioned below in
Sec.~\ref{currentconstraints}.

\subsection{Physical quantities}
\label{physicalquantities}

While the above parameters are useful and conventional 
for characterizing cosmological models, the features in the microwave
background power spectrum depend on various physical quantities which
can be expressed in terms of the parameters. Here the physical
quantities are summarized, and their dependence on parameters given.
This kind of analysis is important for understanding the model
space of parameters as more than just a black box producing output
power spectra. All
of the physical dependences discussed here can be extracted from
Hu and Sugiyama (1996). 
By comparing with numerical solutions to the
evolution equations, Hu and Sugiyama demonstrated that they had
accounted for all relevant physical processes.

Power-law initial conditions are determined in a straightforward way
by the appropriate parameters $Q$, $n$, $r$, and $n_T$, if the
perturbations are purely adiabatic. Additional parameters must be used
to specify any departure from power law spectra, or to specify an
additional admixture of isocurvature initial conditions
(e.g. Bucher {\it et al.} 1999). These
parameters directly express physical quantities.

On the other hand, the physical parameters determining the evolution
of the initial perturbations until decoupling involve a few
specific combinations of cosmological parameters. First, note that the
density of radiation is fixed by the current microwave background
temperature which is known from COBE, as well as the density of the 
neutrino backgrounds. The gravitational
potentials describing scalar perturbations  determine the size of the
Sachs-Wolfe effect and also magnitude of the forces driving the acoustic
oscillations. 
The potentials are determined by $\Omega_0 h^2$, the
matter density as a fraction of critical density. The baryon density,
$\Omega_b h^2$, determines the degree to which the acoustic peak
amplitudes are modulated as described above in Sec.~\ref{baryoneffect}. 

The time of matter-radiation equality is obviously determined solely by the
total matter density $\Omega_0 h^2$. This quantity affects the size of
the dark matter fluctuations, since dark matter starts to collapse
gravitationally only after matter-radiation equality. Also, the
gravitational potentials evolve in time during radiation domination
and not during matter domination: the later matter-radiation equality
occurs, the greater the time evolution of the potentials at
decoupling, increasing the Integrated Sachs-Wolfe effect. 
The power spectrum also has a weak dependence on $\Omega_0$ in
models with $\Omega_0$ significantly less than unity, 
because at late times the evolution of the
background cosmology will be dominated not by matter, but rather 
by vacuum energy (for a flat universe with $\Lambda$) or by
curvature (for an open universe). In either case, the gravitational
potentials once again begin to evolve with time, giving an additional
late-time Integrated Sachs-Wolfe contribution, but this tends to affect
only the largest scales for which the constraints from measurements
are least restrictive due to cosmic variance (see the discussion
in Sec.~\ref{idealexpts} below). 

The sound speed, which sets the sound horizon and thus affects
the wavelength of the acoustic modes (cf. Eq.~(\ref{wkbmodes})),
is completely determined by the baryon density $\Omega_b h^2$. 
The horizon size at recombination, which sets the overall scale of the
acoustic oscillations, depends only on the total mass density
$\Omega_0 h^2$. The damping scale for diffusion damping depends
almost solely on the baryon density $\Omega_b h^2$, although numerical
fits give a slight dependence on $\Omega_b$ alone (Hu and Sugiyama
1996). Finally, the angular diameter distance to the surface of last
scattering is determined by $\Omega_0 h$ and $\Lambda h$; the angular
diameter sets the angular scale on the sky of the acoustic oscillations.

In summary, the physical dependence of the temperature perturbations
at last scattering depends on $\Omega_0 h^2$, $\Omega_b h^2$,
$\Omega_0 h$, and $\Lambda h$ instead of the individual cosmological
parameters $\Omega_0$, $\Omega_b$, $h$, and $\Lambda$. When analyzing
constraints on cosmological models from microwave background power
spectra, it may be more meaningful and powerful to constrain these
physical parameters rather than the cosmological ones.

\subsection{Power spectrum degeneracies}
\label{degeneracies}

As might be expected from the above discussion, not all of the
parameters considered here are independent. In fact, one nearly
exact degeneracy exists if $\Omega_0$, $\Omega_b$, $h$, and $\Lambda$
are taken as independent parameters. To see this, consider a shift
in $\Omega_0$. In isolation, such a shift will produce a corresponding
stretching of the power spectrum in $l$-space. But this effect can be
compensated by first shifting $h$ to keep $\Omega_0 h^2$ constant,
then shifting $\Omega_b$ to keep $\Omega_b h^2$ constant, and finally
shifting $\Lambda$ to keep the angular diameter distance constant.
This set of shifted parameters will, in linear perturbation theory,
produce almost exactly the
same microwave background power spectra as the original set of parameters.
The universe with shifted parameters will generally not be flat, but
the resulting late-time Integrated Sachs-Wolfe effect only weakly
break the degeneracy. Likewise, gravitational lensing has only
a very weak effect on the degeneracy. 

But all is not lost. The required shift in $\Lambda$ is
generally something like 8 times larger than the original shift
in $\Omega_0$, so although the degeneracy is nearly exact,
most of the degenerate models represent rather extreme cosmologies.
Good taste requires either that $\Lambda=0$ or that $\Omega=1$, in
other words that we disfavor models which have both a cosmological
constant and are not flat. If such models are disallowed, the
degeneracy disappears. Finally, other observables not associated with
the microwave background break the degeneracy: the acceleration
parameter $q_0 = \Omega_0/2 - \Lambda$, for example, 
is measured directly by the
high-redshift supernova experiments. So in practice, this
fundamental degeneracy in the microwave background power
spectrum between $\Omega$ and $\Lambda$ is not likely to have
a great impact on our ability to constrain cosmological parameters.

Other approximate degeneracies in the temperature power spectrum
exist between $Q$ and $r$, and between $z_r$ and $n$. The first is 
illusory: the amplitudes of the scalar and tensor power
spectra can be used in place of their sum and ratio, which eliminates
the degeneracy. The power spectrum of large-scale structure will
lift the latter degeneracy if bias is understood well enough, as will
polarization measurements and small-scale second-order temperature
fluctuations (the Ostriker-Vishniac effect, see Jaffe and Gnedin 2000) 
which are both sensitive to $z_r$. 

Finally, many claims have been made about the ability of the microwave
background to constrain the effective number of neutrino species or
neutrino masses. The effective number of massless degrees of freedom
at decoupling can be expressed in terms of the effective number of
neutrino species $N_\nu$ (which does not need to be an integer). This
is a convenient way of parameterizing ignorance about fundamental
particle constituents of nature. Contributors to $N_\nu$ could
include, for example, an extra sterile neutrino sometimes invoked
in neutrino oscillation models, or the thermal background of gravitons
which would exist if inflation did not occur. This parameter can also
include the effects of entropy increases due to decaying or
annihilating particles; see Chapter 3 of Kolb and Turner (1990) for
a detailed discussion. As far as the microwave background is
concerned, $N_\nu$ determines the radiation energy density of the
universe and thus modifies the time of matter-radiation equality.
It can in principle be distinguished from a change
in $\Omega_0 h^2$ because it affects other physical parameters like
the baryon density or the angular diameter distance differently than
a shift in either $\Omega_0$ or $h$. 

Neutrino masses cannot provide
the bulk of the dark matter, because their free streaming greatly
suppresses fluctuation power on galaxy scales, leading to a drastic
mismatch with observed large-scale structure. But models with some
small fraction of dark matter as neutrinos have been advocated to 
improve the agreement between the predicted and observed large-scale
structure power spectrum. Massive neutrinos have several small
effects on the microwave background, which have been studied
systematically by Dodelson {\it et al.}\ (1996). They can slightly
increase the sound horizon at decoupling due to their transition from
relativistic to non-relativistic behavior as the universe expands.
More importantly, free streaming of massive neutrinos around the
time of last scattering leads to a faster decay of the gravitational
potentials, which in turn means more forcing of the acoustic
oscillations and a resulting increase in the monopole perturbations.
Finally, since matter-radiation equality is slightly delayed for
neutrinos with cosmologically interesting masses of a few eV, the
gravitational potentials are less constant 
and a larger Integrated Sachs-Wolfe effect is induced. The change in
sound horizon and shift in matter-radiation equality due to 
massive neutrinos cannot be
distinguished from changes in $\Omega_b h^2$ and $\Omega_0 h^2$, but
the alteration of the gravitational potential's time dependence due to
neutrino free streaming cannot be mimicked  by some other change in
parameters. In principle the effect of neutrino masses can be
extracted from the microwave background, although the effects
are very small.

\subsection{Idealized experiments}
\label{idealexpts}

Remarkably, the microwave background power spectrum
contains enough information to constrain numerous parameters
simultaneously (Jungman {\it et al.} 1996). We would like to estimate
quantitatively just how well the space of parameters described
above can be constrained by ideal measurements of the 
microwave background. The question has been studied in some detail;
this section outlines the basic methods and results, and 
discusses how good various approximations are. For simplicity,
only temperature fluctuations are considered in this section; 
the corresponding formalism for the polarization power spectra
is developed in Kamionkowski {\it et al.}\ (1997).

Given a pixelized map of the microwave sky, we need to determine the
contribution of pixelization noise, detector noise, and beam width to
the multipole moments and power spectrum. 
Consider a temperature map
of the sky $T^{\rm map}({\bf\hat n})$ which is divided into
$N_{\rm pix}$ equal-area pixels. The observed temperature in 
pixel $j$ is due to a
cosmological signal plus noise, $T^{\rm map}_j = T_j + T^{\rm
noise}_j$. 
The multipole coefficients of the map
can be constructed as
\begin{eqnarray}
d^{\rm T}_{lm} &=& {1\over T_0}\int d{\bf\hat n} T^{\rm map}({\bf\hat n}) 
Y_{lm}({\bf\hat n})\cr
&\simeq& {1\over T_0} \sum_{j=1}^{N_{\rm pix}} {4\pi\over N_{\rm pix}}
T^{\rm map}_j Y_{lm}({\bf\hat n}_j),
\label{mapmoments}
\end{eqnarray}
where ${\bf\hat n}_j$ is the direction vector to pixel $j$. 
The map moments are written as $d_{lm}$ to distinguish them from
the moments of the cosmological signal $a_{lm}$; the former include
the effects of noise.  The extent to which the second line in
Eqs.~(\ref{mapmoments}) is only an approximate equality is
the pixelization noise. Most current experiments oversample the sky
with respect to their beam, so the pixelization noise is negligible.
Now assume
that the noise is uncorrelated between pixels and is well-represented
by a normal distribution. Also, assume that the map is created with
a gaussian beam with width $\theta_b$. Then it is straightforward
to show that the variance of the temperature moments is
given by (Knox 1995)
\begin{equation}
\left\langle d^T_{lm} d^{T\, *}_{l'm'}\right\rangle
= \left(C_l e^{-l^2\sigma_b^2} + w^{-1}\right) \delta_{ll'}\delta_{mm'},
\label{dTvariance}
\end{equation}
where $\sigma_b = 0.00742(\theta_b / 1^\circ)$ and
\begin{equation} 
w^{-1} = {4\pi\over N_{\rm pix}}{\left\langle \left(T_i^{\rm noise}\right)^2
\right\rangle \over T_0^2}
\label{winverse}
\end{equation}
is the inverse statistical weight per unit solid angle, a measure of
experimental sensitivity independent of the pixel size.

Now the power spectrum can be 
estimated via Eq.~(\ref{dTvariance}) as
\begin{equation}
C_l^T = (D_l^T - w^{-1})e^{l^2\sigma_b^2}
\label{psestimator}
\end{equation}
where
\begin{equation}
D_l^T = {1\over 2l+1}\sum_{m=-l}^l d^T_{lm} d^{T\, *}_{lm}.
\label{Dldef}
\end{equation}
The individual coefficients $d^T_{lm}$ are gaussian random
variables. This means that $C_l^T$ is a random variable with a
$\chi^2_{2l+1}$ distribution, and its variance is (Knox 1995)
\begin{equation}
\left(\Delta C_l^T\right)^2 = {2\over 2l+1}\left(C_l 
+ w^{-1} e^{l^2\sigma_b^2}\right).
\label{Clvariance}
\end{equation}
Note that even for $w^{-1}=0$, corresponding to zero noise,
the variance is non-zero. This is the cosmic variance, arising
from the fact that we have only one sky to observe: the estimator
in Eq.~(\ref{Dldef}) is the sum of $2l+1$ random variables, so
it has a fundamental
fractional variance of $(2l+1)^{-1/2}$ simply due to Poisson
statistics. 
This variance provides a benchmark for experiments: if the goal is to
determine a power spectrum, it makes no sense to improve resolution or
sensitivity beyond the level at which cosmic variance is the dominant
source of error.

Equation (\ref{Clvariance}) is extremely useful: it gives
an estimate of how well the power spectrum can be determined
by an experiment with a given beam size and detector noise.
If only a portion of the sky is covered, the variance estimate
should be divided by the fraction of the total sky covered.
With these variances in hand, standard statistical techniques
can be employed to estimate how well a given measurement
can recover a given set ${\bf s}$ of cosmological parameters.
Approximate the dependence of $C_l^T$ on a given parameter
as linear in the parameter; this will always be true
for some sufficiently small range of parameter values.
Then the parameter space curvature matrix (also known
as the Fisher information matrix) is specified by 
\begin{equation}
\alpha_{ij} = \sum_l {\partial C_l^T\over \partial s_i}
{\partial C_l^T\over \partial s_j} {1\over(\Delta C_l^T)^2}.
\label{curvaturematrix}
\end{equation}
The variance in the determination of the parameter $s_i$ from
a set of $C_l^T$ with variances $\Delta C_l^T$ after marginalizing
over all other parameters is given by the diagonal element $i$
of the matrix $\alpha^{-1}$. 

Estimates of this kind were first made by Jungman {\it et al.} (1996) and
subsequently refined by Zaldarriaga {\it et al.} (1998) and Bond
{\it et al.} (1998), among others. The basic result is that a map with pixels
of a few arcminutes in size and a signal-to-noise ratio of around 1
per pixel can determine $\Omega$, $\Omega_b h^2$, $\Omega_m
h^2$, $\Lambda h^2$, $Q$, $n$, and $z_r$ at the few percent
level {\it simultaneously}, up to the one degeneracy mentioned above
(see the table in Bond {\it et al.} 1998).  Significant constraints will
also be placed on $r$ and $N_\nu$. This prospect has been the primary
reason that the microwave background has generated such excitement.
Note that $\Omega$, $h$, $\Omega_b$, and $\Lambda$ are the classical
cosmological parameters. Decades of painstaking astronomical
observations have been devoted to determining the values of these
parameters. The microwave background offers a completely independent
method of determining them with comparable or significantly greater
accuracy, and with fewer astrophysical systematic effects to worry
about. The microwave background is also the only source of precise
information about the spectrum and character of the primordial
perturbations from which we arose. Of course, these exciting
possibilities hold only if the universe is accurately represented by a
model in the assumed model space. The model space is, however, quite
broad. Model-independent constraints which the microwave background
provides are discussed in Sec.~\ref{modelindependent}. 

The estimates of parameter variances based on the curvature matrix
would be exact if the power spectrum always varied linearly with
each parameter. This, of course, is not true in general. Given a set
of power spectrum data, we want to know two pieces of information
about the cosmological parameters: (1) What parameter values provide
the best-fit model? (2) What are the error bars on these
parameters, or more precisely, what is the region of parameter space
which defines a given confidence level? The first question can be
answered easily using standard methods of searching parameter space;
generally such a search requires evaluating the power spectrum for
fewer than 100 different models. This shows that the parameter space
is generally without complicated structure or many false minima.
The second question is more
difficult. Anything beyond the curvature matrix analysis
requires looking around in parameter space near the best-fit
model. A specific Monte Carlo technique employing a Metropolis algorithm
has recently been advocated (Christensen and Meyer 2000); such
techniques will certainly prove more flexible and efficient than
recent brute-force grid searches (Tegmark and Zaldarriaga 2000).
As upcoming data sets contain more information and consequently have
greater power to constrain parameters, efficient techniques of
parameter space exploration will become increasingly important.

To this point, the discussion has assumed that the microwave
background power spectrum is perfectly described by linear
perturbation theory. Since the temperature fluctuations are
so small, parts in a hundred thousand, linear theory is a
very good approximation. However, on small scales, non-linear
effects become important and can dominate over the linear
contributions. The most important non-linear effects
are the Ostriker-Vishniac effect coupling velocity and
density perturbations (Jaffe and Kamionkowski 1998, Hu 2000), 
gravitational lensing by large-scale structure 
(Seljak 1996),  
the Sunyaev-Zeldovich effect which gives spectral distortions when
the microwave background radiation passes through hot ionized
regions (Birkinshaw 1999), and the kinetic Sunyaev-Zeldovich effect
which doppler shifts radiation passing through plasma with bulk 
velocity (Gnedin and Jaffe 2000).  
All three effects are measurable and give important
additional constraints on cosmology, but more detailed descriptions
are outside the scope of these lectures.

Finally, no discussion of parameter determination would be complete
without mention of galactic foreground sources of microwave
emission.  
Dust radiates significantly at microwave frequencies, as do
free-free and synchrotron emission; point source microwave emission is
also a potential problem. Dust emission generally
has a spectrum which rises with frequency, while free-free and
synchrotron emission have falling frequency spectra. The emission is
not uniform on the sky, but rather concentrated in the galactic plane,
with fainter but pervasive diffuse emission in other parts of the
sky. The dust and synchrotron/free-free emission spectra cross
each other at a frequency of around 90 GHz. Fortunately for
cosmologists, the amplitude of the foreground emission at this
frequency is low enough to create a frequency window in which the
cosmological temperature fluctuations dominate the foreground
temperature fluctuations. At other frequencies, the foreground contribution
can be effectively separated from the cosmological blackbody
signal by measuring in several different frequencies and
projecting out the portion of the signal with a flat frequency
spectrum. The foreground situation for polarization is less
clear, both in amplitude and spectral index, and could potentially
be a serious systematic limit to the quality of cosmological
polarization data. On the other hand, it may be no greater problem for
polarization fluctuations than for temperature fluctuations. For
an overview of issues surrounding foreground emission, see 
Bouchet and Gispert 1999 or the WOMBAT web site, 
http://astro.berkeley.edu/wombat.

\subsection{Current constraints and upcoming experiments}
\label{currentconstraints}

As the Como School began, results from the high-resolution
balloon-born experiment MAXIMA (Hanany {\it et al.} 2000) were released,
complementing the week-old data from BOOMERanG (de Bernardis {\it et al.} 2000)
and creating a considerable buzz at coffee breaks.  
The derived power spectrum estimates are shown in Fig.~3.
The data from the two measurements appear consistent up to
calibration uncertainties, and
for simplicity will be referred to here as ``balloon data'' and
discussed as a single result. While a few experimenters and data
analyzers were members of both experimental teams, the measurements
and data reductions were done essentially independently.  Earlier data
from the previous year (Miller {\it et al.} 1999) had clearly demonstrated
the existence and angular scale of the first peak in the power
spectrum and produced the first maps of the microwave background at
angular scales below a degree. But the new results from balloon
experiments utilizing extremely sensitive bolometric detecters
represent a qualitative step forward. These experiments begin to
exploit the potential of the microwave background for ``precision
cosmology'';  
their power spectra put strong constraints on several
cosmological parameters simultaneously and rule out many variants of
cosmological models. In fact, what is most interesting is that, at
face value, these measurements put significant pressure on all of the
standard models outlined above.

\begin{figure}
\plotone{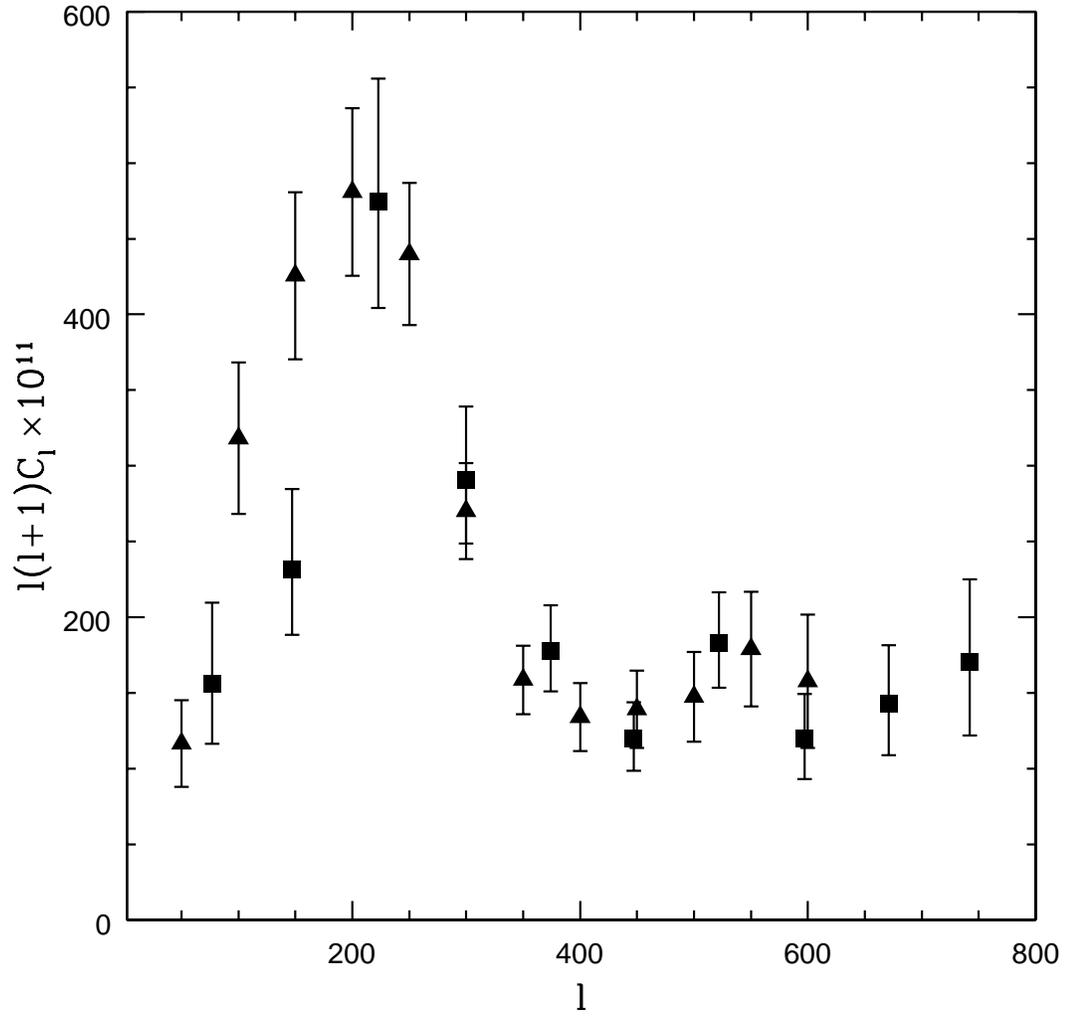}
\caption{Two current measurements of the
microwave background radiation temperature power spectrum. 
Triangles are BOOMERanG measurements
multiplied by 1.21; squares are MAXIMA measurements
multiplied by 0.92. The normalization factors are within
the calibration uncertainties of the experiments, and were
chosen by Hanany et al. (2000) 
to give the most consistent results between the two
experiments. \label{fig3}}
\end{figure}

The balloon data shows two major features: first, a large peak in the
power spectrum centered around $l=200$ with an amplitude of
approximately $l^2C_l = 36000\,\,\mu{\rm K}^2$, and second, a broad plateau
between $l=400$ and $l=700$ with an amplitude of approximately $l^2C_l
= 10000\,\,\mu{\rm K}^2$.
The first peak is clearly delineated and provides good
evidence that the universe is spatially flat, i.e. $\Omega=1$. 
The issue of a second acoustic peak is much less clear. In most flat
universe models with acoustic oscillations, the second peak is
expected to appear at an angular scale of around $l=400$. The angular
resolution of the balloon experiments is certainly good enough to see
such a peak, but the power spectrum data shows no evidence for one. I
argue that a flat line is an excellent fit to the data past $l=300$,
and that any model which shows a peak in this region will be a worse
fit than a flat line.  This does not necessarily mean that no peak is
present; the error bars are too large to rule out a peak, but the
amplitude of such a peak is fairly strongly constrained to be lower
than expected given the first peak.

What does this mean for cosmological models? Within the model
space outlined in the previous section, there are
three ways to suppress the second peak. The first would be to have a
power spectrum index $n$ substantially less than 1. This solution
would force abandonment of the assumption of power law initial
fluctuations, in order to match the observed amplitude of large-scale
structure at smaller scales. While this is certainly possible, it
represents a drastic weakening in the predictive power of the
microwave background: essentially, a certain feature is
reproduced by arbitrarily changing the primordial power
spectrum. While no physical principle requires power law primordial
perturbations, we should wait for microwave background measurements on
a much wider range of scales combined with overlapping large-scale
structure measurements before resorting to departures from power-law
initial conditions. If the universe really did possess an initial
power spectrum with a variety of features in it, most of the promise
of precision cosmology is lost. Recent power spectra extracted from
the IRAS Point Source Survey Redshift Catalog (Tegmark and Hamilton
2000), which show a remarkable power law behavior spanning three
orders of magnitude in wave number, seem to argue against this
possibility.

The second possibility is a drastic amount of reionization. It is not
clear the extent to which this might be compatible with the height of
the first peak and still suppress the second peak sufficiently. This
possibility seems unlikely as well, but would show clear signatures in
the microwave background polarization.

The most commonly discussed possibility is that the very low second
peak amplitude reflects an unexpectedly large fraction of baryons
relative to dark matter in the universe. The baryon signature
discussed in Sec.~\ref{baryoneffect} 
gives a suppression of the second peak in
this case. 
However, primordial nucleosynthesis also constrains the
baryon-photon ratio. Recent high-precision measurements of deuterium
absorption in high-redshift neutral hydrogen clouds (Tytler {\it et al.} 2000) give
a baryon-photon number ratio of $\eta=5.1\pm 0.5 \times 10^{10}$, 
which translates to
$\Omega_b h^2 = 0.019\pm0.002$ assuming that the entropy (i.e. photon
number) per comoving volume remains constant between nucleosynthesis
and the present. 
Requiring $\Omega_b$ to satisfy this nucleosynthesis
constraint leads to microwave background power spectra which are not
particularly good fits to the data. An alternative is that the entropy
per comoving volume has {\it not} remained fixed between
nucleosynthesis and recombination (see, e.g., Kaplinghat and Turner
2000). This could be arranged by having a
dark matter particle which decays to photons, although such a process
must evade limits from the lack of microwave background spectral
distortions (Hu and Silk 1993). Alternately, a
large chemical potential for the neutrino background could lead to
larger inferred values for the baryon-photon ratio from
nucleosynthesis (Esposito {\it et al.} 2000).  
Either way, if both the microwave background
measurements and the high-redshift deuterium abundances hold up, the
discrepancy points to new physics.
Of course, a final explanation for the discrepancies is simply that
the balloon data has significant systematic errors.

I digress for a brief editorial comment about data analysis.
Straightforward searches of the conventional cosmological model space
described above for good fits to the balloon data give models with
very low dark matter densities, high baryon fractions, and very large
cosmological constants 
(see model P1 in Table 1 of Lange {\it et al.} 2000). Such models
violate other observational constraints on age, which must be at least
12 billion years (see, e.g., Peacock {\it et al.} 1998), and quasar and radio
source strong
lensing number counts, which limit a cosmological constant to $\Lambda
\leq 0.7$ (Falco {\it et al.} 1998). The response to this situation so far has
been to invoke Bayesian prior probability distributions on various
quantities like $\Omega_b$ and the age. This leads to a best-fit
model with a nominally acceptable $\chi^2$ (Lange {\it et al.} 2000, Tegmark
{\it et al.} 2000 and others). 
But be wary of this procedure when the priors
have a large effect on the best-fit model! The microwave background
will soon provide tighter constraints on most parameters than any
other source of prior information. Priors probabilities on a given
parameter are useful and justified when the microwave background data
has little power to constrain that parameter; in this case, the
statistical quality of the model fit to the microwave background data
will not be greatly affected by imposing the prior. However, something
fishy is probably going on when a prior pulls a parameter multiple
sigma away from its best fit value without the prior. This is what
happens presently with $\Omega_b$ when the nucleosynthesis prior is
enforced. If your priors make a big difference, it is likely either
that some of the data is incorrect, or that the model space does not
include the correct model. Both the microwave background measurements
and the high-redshift deuterium detections are taxing observations
dominated by systematic effects, so it is certainly possible that one
or both are wrong. On the other hand, MAXIMA and BOOMERanG are
consistent with each other while using different instruments,
different parts of the sky, and different analysis pipelines, and the
deuterium measurements are consistent for several different
clouds. This suggests possible missing physics ingredients, like
extreme reionization or an entropy increase mentioned above, or
perhaps significant contributions from
cosmic defects.  It has even been suggested by
otherwise sober and reasonable people that the microwave background
results, combined with various difficulties related to dynamics of
spiral galaxies, may point towards a radical revision of the standard
cosmology (Sellwood and Kosowsky 2000).  We should not rest lightly
until the cosmological model preferred by microwave background
measurements is comfortably consistent with all relevant priors
derived from other data sources of comparable precision.

The picture will come into sharper relief over the next two years.
The MAP satellite (http://map.gsfc.nasa.gov), due for launch by NASA
in May, 2001, will map the full microwave sky in five frequency
channels with an angular resolution of around 15 arc minutes and a
temperature sensitivity per pixel of a part in a million. Space
missions offer unequalled sky coverage and control of systematics, and
if it works as advertised, MAP will be a benchmark experiment. Prior
to its launch, expect to see the first interferometric microwave data
at angular scales smaller than a half degree from the CBI 
interferometer experiment (http://www.astro.caltech.edu/\~{}tjp/CBI/). 
In this same time frame,
we also may have the first detection of polarization.  The most
interesting power spectrum feature to focus on will be the existence
and amplitude of a third acoustic peak. If a third peak appears with
amplitude significantly higher than the putative second peak, this
almost certainly indicates conventional acoustic oscillations with a
high baryon fraction and possibly new physics to reconcile the result
with the deuterium measurements. If, on the other hand, the power
spectrum remains flat or falls further past the second peak region,
then all bets are off. In a time frame of the next 5 to 10 years,
we can reasonably expect to have a cosmic-variance limited temperature
power spectrum down to scales of a few arcminutes (say, $l=4000$),
along with significant polarization information (though probably
not cosmic-variance limited power spectra). In particular, ESA's
Planck satellite mission 
(http://astro.estec.esa.nl/SA-general/Projects/Planck/) 
will map the microwave sky in
nine frequency bands at significantly better resolution and
sensitivity than the MAP mission. For a
comprehensive listing of past and planned microwave background
measurements, see Max Tegmark's experiments web page,\hfil\break
http://www.hep.upenn.edu/\~{}max/cmb/experiments.html.

\section{Model-Independent Cosmological Constraints}
\label{modelindependent}

Most analysis of microwave background data and predictions
about its ability to constrain cosmology have been based on
the cosmological parameter space described in 
Sec.~\ref{spaceofmodels} above. This space is motivated
by inflationary cosmological scenarios, which generically
predict power-law adiabatic perturbations evolving only via
gravitational instability. Considering that this
space of models is broad and appears to fit all current
data far better than any other proposed models, such an
assumed model space is not very restrictive. In particular, proposed
extensions tend to be rather ad hoc, adding
extra elements to the model without providing any compelling
underlying motivation for them. Examples which have been
discussed in the literature include multiple types of dark matter
with various properties, nonstandard recombination, small
admixtures of topological defects, production of excess entropy,
or arbitrary initial power spectra. None of these possibilities
are attractive from an aesthetic point of view: all add significant
complexity and freedom to the models without any corresponding
restrictions on the original parameter space. The principle of
Occam's Razor should cause us to be skeptical about any such
additions to the space of models. 

On the other hand, it is possible that some element {\it is} missing
from the model space, or that the actual cosmological model is
radically different in some respect. The microwave background is the
probe of cosmology most tightly connected to the fundamental
properties of the universe and least influenced by astrophysical
complications, and thus the most capable data source for deciding
whether the universe actually is well described by some model in
the usual model space. An interesting question is the extent to which
the microwave background can determine various properties of the
universe independent from particular models.  While any
cosmological interpretation of temperature fluctuations in the
microwave sky requires some kind of minimal assumptions, all of
the conclusions outlined below can be drawn without invoking a
detailed model of initial conditions or structure formation.
These conclusions are in contrast to precision determination of
cosmological parameters, which does require the assumption of a
particular space of models and which can vary significantly depending
on the space.

\subsection{Flatness}
\label{flatnesstests}

The Friedmann-Robertson-Walker spacetime describing homogeneous and
isotropic cosmology comes in three flavors of spatial curvature:
positive, negative, and flat, corresponding to $\Omega > 1$,
$\Omega < 1$, and $\Omega = 1$ respectively. One of the most
fundamental questions of cosmology, dating to the original
relativistic cosmological models, is the curvature of the background
spacetime. The fate of the universe quite literally depends on the
answer: in a cosmology with only matter and radiation, a positively-curved
universe will eventually recollapse in a fiery ``Big Crunch'' while flat and
negatively-curved universes will expand forever, meeting a frigid
demise. Note these fates are at least 40 billion years in the future. (A
cosmological constant or other energy density component with an
unusual equation of state can alter these outcomes, causing a closed
universe eventually to enter an inflationary stage.)

The microwave background provides the cleanest and most powerful probe
of the geometry of the universe (Kamionkowski {\it et al.}\ 1994). 
The surface of last scattering is at
a high enough redshift that photon geodesics between the last
scattering surface and the Earth are significantly curved if the
geometry of the universe is appreciably different than flat. In a
positively-curved space, two geodesics will bend towards each other,
subtending a larger angle at the observer than in the flat case;
likewise, in a negatively-curved space two geodesics bend away from
each other, resulting in a smaller observed angle between the two. 
The operative quantity is the angular diameter distance; Weinberg
(2000) gives a pedagogical discussion of its dependence on
$\Omega$. In a flat universe, the horizon length at the time of last
scattering subtends an angle on the sky of around two degrees. For a
low-density universe with $\Omega = 0.3$, this angle becomes smaller
by half, roughly. 

A change in angular scale of this magnitude will change the apparent
scale of all physical scales in the microwave background. A
model-independent determination of $\Omega$ thus requires a physical
scale of known size to be imprinted on the primordial plasma at last
scattering; this physical scale can then be compared with its apparent
observed scale to obtain a measurement of $\Omega$. The microwave
background fluctuations actually depend on two basic physical scales.
The first is the sound horizon at last scattering, $r_s$
(cf. Eq.~(\ref{soundhorizon}).  
If coherent acoustic oscillations are
visible, this scale sets their characteristic wavelengths. Even if
coherent acoustic oscillations are not present, the sound horizon
represents the largest scale on which any causal physical process can
influence the primordial plasma. Roughly, if primordial perturbations
appear on all scales, the resulting microwave background fluctuations appear
as a featureless power law at large scales, while the scale at which they
begin to depart from this assumed primordial behavior corresponds to
the sound horizon. This is precisely the behavior observed by current
measurements, which show a prominent power spectrum peak at an angular
scale of a degree ($l=200$), arguing strongly for a flat universe. 
Of course, it is logically possible that the primordial power
spectrum has power on scales only significantly smaller than the
horizon at last scattering. In this case, the largest scale
perturbations would appear at smaller angular scales for a given
geometry. But then the observed power-law perturbations at large
angular scales must be reproduced by the Integrated Sachs-Wolfe
effect, and resulting models are contrived.
If the microwave background power spectrum exhibites acoustic 
oscillations, then the spacing of the acoustic peaks depends
only on the sound horizon independent of the phase of the
oscillations; this provides a more general and precise probe of
flatness than the first peak position.

The second physical scale provides another test: the Silk
damping scale is determined solely by the thickness of the surface
of last scattering, which in turn depends only on the baryon density
$\Omega_b h^2$, the expansion rate of the universe and 
standard thermodynamics. Observation of an 
exponential suppression of power at small scales gives an estimate
of the angular scale corresponding to the damping scale. Note that the
effects of reionization and gravitational lensing must both be
accounted for in the small-scale dependence of the fluctuations. If
the reionization redshift can be accurately estimated from microwave
background polarization (see below) and the baryon density is known
from primordial nucleosynthesis or from the alternating peak heights
signature (Sec.~\ref{baryoneffect}), only a radical modification of the
standard cosmology altering the time dependence of the scale factor or
modifying thermodynamic recombination can change the physical 
damping scale. If the estimates of $\Omega$
based on the sound horizon and damping scales are consistent,
this is a strong indication that the inferred geometry of the universe
is correct.

\subsection{Coherent acoustic oscillations}
\label{coherenttests}

If a series of peaks equally spaced in $l$ is observed in the
microwave background temperature power spectrum, it strongly suggests
we are seeing the effects of coherent acoustic oscillations at the
time of last scattering. Microwave background polarization provides a
method for confirming this hypothesis. 
As explained in Sec.~\ref{thomsonscattering},
polarization anisotropies couple primarily to velocity perturbations,
while temperature anisotropies couple primarily to density
perturbations. Now coherent acoustic oscillations produce temperature
power spectrum peaks at scales where a mode of that wavelength has
either maximum or minimum compression in potential wells at the time
of last scattering.  The fluid velocity for the mode at these times
will be zero, as the oscillation is turing around from expansion to
contraction (envision a mass on a spring.)  At scales intermediate
between the peaks, the oscillating mode has zero density contrast but
a maximum velocity perturbation.  Since the polarization power
spectrum is dominated by the velocity perturbations, its peaks will be
at scales interleaved with the temperature power spectrum peaks. This
alternation of temperature and polarization peaks as the angular scale
changes is characteristic of acoustic oscillations (see Kosowsky
(1999) for a more detailed discussion).  Indeed, it is almost like
seeing the oscillations directly: it is difficult to imagine any other
explanation for density and velocity extrema on alternating scales.
The temperature-polarization cross-correlation must also have peaks
with corresponding phases.  This test will be very useful if a series
of peaks is detected in a temperature power spectrum which is not a
good fit to the standard space of cosmological models. If the peaks
turn out to reflect coherent oscillations, we must then modify some
aspect of the underlying cosmology, while if the peaks are not
coherent oscillations, we must modify the process by which
perturbations evolve.

If coherent oscillations are detected, any cosmological model must
include a mechanism for enforcing coherence. Perturbations on all
scales, in particular on scales outside the horizon, provide the only
natural mechanism: the phase of the oscillations is determined by the
time when the wavelength of the perturbation becomes smaller than the
horizon, and this will clearly be the same for all perturbations of a
given wavelength. For any source of perturbations inside the horizon,
the source itself must be coherent over a given scale to produce
phase-coherent perturbations on that scale. This cannot occur without
artificial fine-tuning. 

\subsection{Adiabatic primordial perturbations}
\label{adiabatictests}

If the microwave background temperature and polarization power spectra
reveal coherent acoustic oscillations and the geometry of the universe
can also be determined with some precision, then the phases of the
acoustic oscillations can be used to determine whether the primordial
perturbations are adiabatic or isocurvature. Quite generally,
Eq.~(\ref{wkbmodes}) shows that adiabatic and isocurvature
power spectra must have peaks which are out of phase. While current
measurements of the microwave background and large-scale structure
rule out models based entirely on isocurvature perturbations, some
relatively small admixture of isocurvature modes with dominant
adiabatic modes is possible. Such mixtures arise naturally in
inflationary models with more than one dynamical field during
inflation (see, e.g., Mukhanov and Steinhardt 1998).

\subsection{Gaussian primordial perturbations}
\label{gaussiantests}

If the temperature perturbations are well approximated as
a gaussian random field, as microwave background maps so far suggest,
then the power spectrum $C_l$ contains all statistical information about the
temperature distribution. Departures from gaussianity take myriad
different forms; the business of providing general but useful
statistical descriptions is a complicated one (see, e.g., Ferreira 
{\it et al.} 1997). Tiny amounts of nongaussianity will arise
inevitably from non-linear evolution of fluctuations, and larger
nongaussian contributions can be a feature of the primordial
perturbations or can be induced by ``stiff'' stress-energy
perturbations such as topological defects. As explained below,
defect theories of structure formation seem to be ruled out
by current microwave background and large-scale structure
measurements, so interest in nongaussianity has waned. But the extent
to which the temperature fluctuations are actually gaussian is
experimentally answerable, and as observations improve this will
become an important test of inflationary cosmological models.

\subsection{Tensor or vector perturbations}
\label{tensortests}

As described in Sec.~\ref{harmonicexpansions}, 
the tensor field describing microwave
background polarization can be decomposed into two components
corresponding to the gradient-curl decomposition of a vector field.
This decomposition has the same physical meaning as that for a vector
field. In particular, any gradient-type tensor field, composed of
the G-harmonics, has no curl,
and thus may not have any handedness associated with it (meaning the
field is even under
parity reversal), while the curl-type tensor field, composed of the
C-harmonics, does have a handedness (odd under parity reversal).

This geometric interpretation leads to an important physical
conclusion. Consider a universe containing only scalar perturbations,
and imagine a single Fourier mode of the perturbations. The mode has
only one direction associated with it, defined by the Fourier vector
${\bf k}$; since the perturbation is scalar, it must be rotationally
symmetric around this axis.  (If it were not, the gradient of the
perturbation would define an independent physical direction, which
would violate the assumption of a scalar perturbation.)  Such a mode
can have no physical handedness associated with it, and as a result,
the polarization pattern it induces in the microwave background
couples only to the G harmonics. Another way of stating this
conclusion is that primordial density perturbations produce {\it no}
C-type polarization as long as the perturbations evolve linearly.
On the other hand, primordial tensor or vector perturbations produce both
G-type and C-type polarization of the microwave background
(provided that the tensor or vector perturbations
themselves have no intrinsic net polarization associated with them).

Measurements of cosmological C-polarization in the microwave
background are free of contributions from the dominant scalar
density perturbations and thus can reveal the contribution of tensor
modes in detail. For roughly scale-invariant tensor perturbations,
most of the contribution comes at angular scales larger than
$2^\circ$ ($2 < l < 100$). Figure 4 displays the C and G power
spectra for scale-invariant tensor perturbations contributing
10\% of the COBE signal on large scales. A microwave background map with
forseeable sensitivity could measure gravitational wave
perturbations with amplitudes smaller than $10^{-3}$ times
the amplitude of density perturbations (Kamionkowski and Kosowsky
1998). The C-polarization signal also appears to be the best
hope for measuring the spectral index $n_T$ of the tensor perturbations. 

\begin{figure}
\plotone{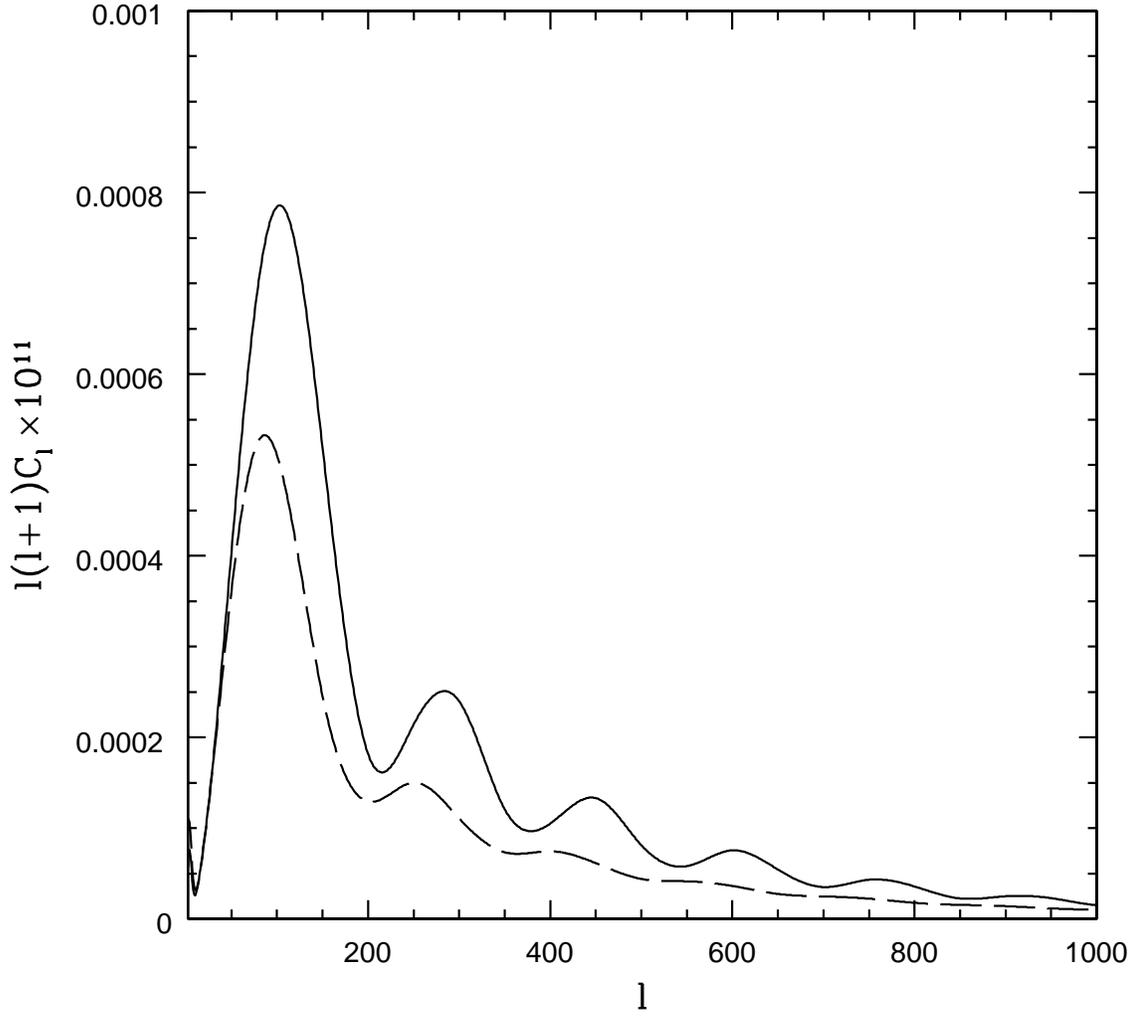}
\caption{Polarization power spectra from tensor perturbations: the solid
line is $C_l^{\rm G}$ and the dashed line is $C_l^{\rm C}$. The amplitude
gives a 10\% contribution to the COBE temperature power spectrum 
measurement at low l. Note
that scalar perturbations give no contribution to $C_l^{\rm C}$.
\label{fig4}}
\end{figure}

\subsection{Reionization redshift}
\label{reionizationtests}

Reionization produces a distinctive microwave background signature.
It suppresses temperature fluctuations by increasing
the effective damping scale, while it also increases large-angle polarization
due to additional Thomson scattering at low redshifts when the
radiation quadrupole fluctuations are much larger. This enhanced
polarization peak at large angles will be significant for reionization
prior to $z=10$ (Zaldarriaga 1997). Reionization will also greatly enhance the
Ostriker-Vishniac effect, a second-order coupling between density and
velocity perturbations (Jaffe and Kamionkowski 1998).
The nonuniform reionization inevitable if the ionizing photons come
from point sources, as seems likely, may also create an additional
feature at small angular scales (Hu and Gruzinov 1998, Knox {\it et al.} 1998). Taken
together, these features are clear indicators of the reionization
redshift $z_r$ independent of any cosmological model.

\subsection{Magnetic Fields}
\label{bfieldstests}

Primordial magnetic fields would be clearly indicated if cosmological
Faraday rotation were detected in the microwave background
polarization. A field with comoving field strength of $10^{-9}$ gauss
would produce a signal with a few degrees of rotation at 30 GHz, which
is likely just detectable with future polarization experiments 
(Kosowsky and Loeb 1996). 
Faraday rotation has the effect of mixing G-type and
C-type polarization, and would be another contributor to the
C-polarization signal, along with tensor perturbations. Depolarization
will also result from Faraday rotation in the case of significant
rotation through the last scattering surface (Harari {\it et al.}\ 1996)
Additionally, the tensor and vector metric perturbations produced by
magnetic fields result in further microwave background fluctuations.
A distinctive signature of such fields is that for a range of
power spectra, the polarization fluctuations from the metric perturbations
is comparable to, or larger than, the corresponding
temperature fluctuations
(Kahniashvili {\it et al.} 2000). Since the microwave  background power spectra
vary as the fourth power of the magnetic field amplitude, it is
unlikely that we can detect magnetic fields with comoving amplitudes
significantly below $10^{-9}$ gauss. However, if such fields do exist,
the microwave background provides several correlated signatures which
will clearly reveal them.

\subsection{The topology of the universe}
\label{topologytests}

Finally, one other microwave background signature of a very different
character deserves mention. Most cosmological analyses make the
implicit assumption that the spatial extent of the universe is
infinite, or in practical terms at least much larger than our current
Hubble volume so that we have no way of detecting the bounds of the
universe. However, this need not be the case. The requirement that the
unperturbed universe be homogeneous and isotropic determines the
spacetime metric to be of the standard Friedmann-Robertson-Walker
form, but this is only a {\it local} condition on the spacetime.
Its global structure is still unspecified. It is possible to construct
spacetimes which at every point have the usual homogeneous and
isotropic metric, but which are spatially compact (have finite
volumes). The most familiar example is the construction of a
three-torus from a cubical piece of the flat spacetime by
identifying opposite sides. Classifying the possible topological
spaces which locally have the metric structure of the usual
cosmological spacetimes (i.e. have the Friedmann-Robertson-Walker
spacetimes as a topological covering space) has been studied
extensively. The zero-curvature and positive-curvature
cases have only a handful of possible topological spaces
associated with them, while the negative curvature case has
an infinite number with a very rich classification. See
Weeks (1998) for a review.

If the topology of the universe is non-trivial and the volume of the
universe is smaller than the volume contained by a sphere with radius
equal to the distance to the surface of last scattering, then it is
possible to detect the topology. Cornish {\it et al.} (1996) pointed
out that because the last scattering surface is always a sphere in the
covering space, any small topology will result in matched circles of
temperature on the microwave sky. The two circles represent photons
originating from the same physical location in the universe but
propagating to us in two different directions. Of course, the
temperatures around the circles will not match exactly, but only the
contributions coming from the Sachs-Wolfe effect and the intrinsic
temperature fluctuations will be the same; the velocity and Integrated
Sachs-Wolfe contributions will differ and constitute a noise source.
Estimates show the circles can be found efficiently
via a direct search of full-sky microwave background maps.
Once all matching pairs of circles have been discovered, their number
and relative locations on the sky strongly overdetermine the topology
of the universe in most cases. Remarkably, the microwave background
essentially allows us to determine the size of the universe if it is
smaller than the current horizon volume in any dimension. 

\section{Finale: Testing Inflationary Cosmology}
\label{finale}

In summary, the cosmic microwave background radiation is a remarkably
interesting and powerful source of information about cosmology. It
provides an image of the universe at an early time when the
relevant physical processes were all very simple, so the dependence of
anisotropies on the cosmological model can be calculated with high
precision. At the same time, the universe at decoupling was an
interesting enough place that small differences in cosmology will
produce measurable differences in the anisotropies.

The microwave background has the ultimate potential to 
determine fundamental cosmological parameters describing the universe
with percent-level precision. If this promise is realized, the
standard model of cosmology would compare with the standard model
of particle physics in terms of physical scope, explanatory power,
and detail of confirmation. But in order for such a situation to
come about, we must first choose a model space which includes
the correct model for the universe. The accuracy with which
cosmological parameters can be determined is of course limited by the
accuracy with which some model in the model space represents the
actual universe.  

The space of models discussed in Sec.~\ref{spaceofmodels} represents universes
which we would expect to arise from the mechanism of inflation. These
models have become the standard testing ground for comparisons with
data because they are simple, general, and well-motivated. So far,
these types of models fit the data well, much better than any
competing theories.  Future measurements may remain perfectly
consistent with inflationary models, may reveal inconsistencies which
can be remedied via minor extensions or modifications of the parameter
space, or may require more serious departures from these types of
models.

For the sake of a concluding discussion about the power of the
microwave background, assume that the universe actually is well
described by inflationary cosmology, and that it can be modelled by
the parameters in Sec.~\ref{spaceofmodels}. 
For an overview of inflation and the
problems it solves, see Kolb and Turner (1990, chapter 8) or the
lectures of A.~Linde in this volume.  
To what extent can we hope to
verify inflation, a process which likely would have occurred at an
energy scale of $10^{16}$ GeV when the universe was $10^{-38}$ seconds
old? Direct tests of physics at these energy scales are unimaginable,
leaving cosmology as the only likely way to probe this physics.

Inflation is not a precise theory, but rather a mechanism for
exponential expansion of the universe which can be realized in a
variety of specific physical models. Cosmology in general and the
cosmic microwave background in particular can hope to test the
following predictions of inflation (see Kamionkowski and Kosowsky 1999
for a more complete discussion of inflation and its observable
microwave background properties):

\begin{itemize}

\item{} The most basic prediction of inflation is a spatially flat
universe. The flatness problem was one of the fundamental motivations
for considering inflation in the first place. While it is possible to
construct models of inflation which result in a non-flat universe,
they all must be finely tuned for inflation to end at just the right
time for a tiny but non-negligible amount of curvature to remain. The
geometry of the universe is one of the fundamental pieces of physics
which can be extracted from the microwave background power
spectra. Recent measurements make a strong case that the universe is
indeed flat. 

\item{} Inflation generically predicts primordial perturbations which
have a gaussian statistical distribution. The microwave background is
the only precision test of this prediction. Primordial gaussian
perturbations will still be almost precisely gaussian at
recombination, whereas they will have evolved significant
nongaussianity by the time the local large-scale structure forms, due
to gravitational collapse. Other methods of probing gaussianity, like
number densities of galaxies or other objects, inevitably depend
significantly on astrophysical modelling.

\item{} The simplest models of inflation, with a single dynamical
scalar field, give adiabatic primordial perturbations. The only real
test of this prediction comes from the microwave background power
spectrum.  More complex models of inflation with multiple dynamical
fields generically result in dominant adiabatic fluctuations with some
admixture of isocurvature fluctuations. Limits on isocurvature
fluctuations obtained from microwave background measurements could be
used to place constraints on the size of couplings between different
fields at inflationary energy scales.

\item{} Inflation generically predicts primordial perturbations
on all scales, including scales outside the horizon. Of course
we can never test directly whether perturbations on scales larger than
the horizon exist, but the microwave background can reveal
perturbations at recombination on scales comparable to the horizon
scale. Zaldarriaga and Spergel (1997) have argued that inflation
generically gives a peak in the polarization power spectrum at
angular scales larger than $2^\circ$, and that no causal perturbations
at the epoch of last scattering can produce a feature at such large
scales. Inflation further predicts that the primordial power spectrum
should be close to a scale-invariant power law (e.g. Huterer and
Turner 2000), although complicated models can lead to
power spectra with features or significant departures from 
scale invariance. The microwave background can probe the primordial
power spectrum over three orders of magnitude. 

\item{} Inflationary perturbations result in phase-coherent
acoustic oscillations. The coherence arises because on any given
scale, the perturbations start in the same state determined only
by their character outside the horizon. For a discussion in
the language of squeezed quantum states, see Albrecht (2000). 
It is extremely difficult to produce coherent oscillations by any
mechanism other than perturbations outside the horizon. The microwave
background temperature and polarization power spectra will together
clearly reveal coherent oscillations. 

\item{} Inflation finally predicts potentially measurable relationships
between the amplitudes and power law indices of the primordial density
and gravitational wave perturbations (see Lidsey {\it et al.} 1997 for a
comprehensive overview), and measuring a $C_l^{\rm C}$ power spectrum
appears to be the only way to obtain precise enough measurements of
the tensor perturbations to test these predictions, thanks to the
fact that the density perturbations don't contribute to $C_l^{\rm C}$.
Detection of inflationary tensor perturbations would reveal the energy
scale at which inflation occurred, while confirming the inflationary
relationships between scalar and tensor perturbations would provide a
strong consistency check on inflation. 

\end{itemize}

The potential power of the microwave background is demonstrated by the
fact that inflation, a theoretical mechanism which likely would occur at
energy scales not too different from the Planck scale, would result
in several distinctive signatures in the microwave background
radiation. Current measurements beautifully confirm a flat universe
and are fully consistent with gaussian perturbations; the rest of the
tests will come into clearer view over the coming years. If inflation
actually occurred, we can expect to have very strong circumstantial
supporting evidence from the above signatures, 
along with precision measurements of the
cosmological parameters describing our universe. 
On the other hand, if inflation did not occur, the universe will
likely look different in some respects from the space of models
in Sec.~\ref{spaceofmodels}. In this case, we may not be able to recover
cosmological parameters as precisely, but the microwave background
will be equally important in discovering the correct model of our universe.

\acknowledgments

I thank the organizers for a stimulating and enjoyable Summer School.
The preparation of these lectures has been supported by the
NASA Astrophysics Theory Program and the
Cotrell Scholars program of the Research Corporation.

\end{document}